\documentclass[12pt]{article}
\usepackage{latexsym,amsmath,amssymb,graphicx}
\renewcommand{\thefootnote}{\fnsymbol{footnote}}

\usepackage{amsmath}
\usepackage{amsfonts}

\numberwithin{equation}{section}

\def\doubleset#1#2{\bgroup%
\def\doit#1#2{%
\setbox\dblsetbox=\hbox{$\cstyle #1$}%
\raise#2\ht\dblsetbox\copy\dblsetbox%
\hskip-\wd\dblsetbox%
\raise-#2\ht\dblsetbox\box\dblsetbox}%
\mathchoice%
{\def\cstyle{\displaystyle}\doit#1#2}%
{\def\cstyle{\textstyle}\doit#1#2}%
{\def\cstyle{\scriptstyle}\doit#1#2}%
{\def\cstyle{\scriptscriptstyle}\doit#1#2}\egroup}
\def\underarrow#1{\vbox{\ialign{##\crcr$\hfil\displaystyle
 {#1}\hfil$\crcr\noalign{\kern1pt\nointerlineskip}$\longrightarrow$\crcr}}}

\newbox\dblsetbox

\newlength{\extraspace}
\newlength{\extraspaces}
\newcommand{\be}{\begin{equation}
\addtolength{\abovedisplayskip}{\extraspaces}
\addtolength{\belowdisplayskip}{\extraspaces}
\addtolength{\abovedisplayshortskip}{\extraspace}
\addtolength{\belowdisplayshortskip}{\extraspace}}
\newcommand{\ee}{\end{equation}}

\newcommand{\ba}{\begin{eqnarray}
\addtolength{\abovedisplayskip}{\extraspaces}
\addtolength{\belowdisplayskip}{\extraspaces}
\addtolength{\abovedisplayshortskip}{\extraspace}
\addtolength{\belowdisplayshortskip}{\extraspace}}
\newcommand{\ea}{\end{eqnarray}}

\newcommand{\bd}{\begin{displaymath}
\addtolength{\abovedisplayskip}{\extraspaces}
\addtolength{\belowdisplayskip}{\extraspaces}
\addtolength{\abovedisplayshortskip}{\extraspace}
\addtolength{\belowdisplayshortskip}{\extraspace}}
\newcommand{\ed}{\end{displaymath}}

\newcounter{saveeqn}

\newcommand{\newsection}[1]{
\vspace{12mm} \pagebreak[3] \addtocounter{section}{1}
\setcounter{equation}{0} \setcounter{subsection}{0}
\noindent{\bf \thesection. #1} \nopagebreak
\medskip
\nopagebreak
\addcontentsline{toc}{section}{\thesection. #1}}

\newcommand{\newsubsection}[1]{
\vspace{0.8cm} \pagebreak[3] \addtocounter{subsection}{1}
\setcounter{subsubsection}{0}
\noindent{ \it \thesubsection. #1} \nopagebreak \vspace{2mm}
\nopagebreak
\addcontentsline{toc}{subsection}{\thesubsection. #1}}

\begin{document}
\addtolength{\baselineskip}{1.5mm}

\thispagestyle{empty}

\vbox{} \vspace{2.0cm}

\begin{center}
\centerline{\LARGE{Surface Operators in ${\cal N}=2$ Abelian Gauge Theory}}
\bigskip

\vspace{2.0cm}

{Meng-Chwan~Tan \footnote{On leave of absence from the National
University of Singapore.}}
\\[2mm]
{\it California Institute of Technology, \\
Pasadena, CA 91125, USA} \\ [1mm]
e-mail: mengchwan@theory.caltech.edu\\
\end{center}

\vspace{2.0 cm}

\centerline{\bf Abstract}\smallskip \noindent

  We generalise the analysis in [arXiv:0904.1744] to superspace, and explicitly prove that for any embedding of surface operators in a general, twisted ${\cal N} =2$ pure abelian theory on an arbitrary spin (or non-spin) four-manifold, the parameters transform naturally under the $SL(2,\mathbb Z)$ (or $\Gamma_0(2)$) duality of the theory. However, for nontrivially-embedded surface operators, exact $S$-duality holds if and only if the ``quantum'' parameter effectively vanishes, while the overall $SL(2,\mathbb Z)$ (or $\Gamma_0(2)$) duality holds up to a $c$-number at most, regardless. Nevertheless, this observation sets the stage for a physical proof of a remarkable mathematical result by Kronheimer and Mrowka~\cite{KM}---that expresses a ``ramified'' analog of the Donaldson invariants solely in terms of the ordinary Donaldson invariants---which, will appear, among other things, in forthcoming work~\cite{MC}. As a prelude to that, the effective interaction on the corresponding $u$-plane will be computed. In addition, the dependence on second Stiefel-Whitney classes and the appearance of a $\textrm{Spin}^c$ structure in the associated low-energy Seiberg-Witten theory with surface operators, will also be demonstrated. In the process, we will stumble upon an interesting phase factor that is otherwise  absent in the ``unramified'' case.

 \newpage

\renewcommand{\thefootnote}{\arabic{footnote}}
\setcounter{footnote}{0}

\tableofcontents

%\newpage

\vspace{0.0cm}

\newsection{Introduction and Summary}

Surface operators are higher-dimensional analogues of the usual Wilson and 't Hooft loop operators in gauge theory that are supported on a codimension two submanifold of spacetime. They are defined by specifying a certain type of singularity in the relevant fields as one approaches the submanifold. Such operators were first used to probe the dynamics of gauge theory and black holes in~\cite{Preskill}-\cite{Bucher}. Thereafter, they appeared in the mathematical literature in an application to Donaldson theory~\cite{KM, Mrowka}, and in the relation between instantons, Seiberg-Witten theory and integrable systems~\cite{Braverman, Etingof}.

More recently, in an effort to furnish a gauge-theoretic interpretation of the geometric Langlands program with ramification, surface operators have also been considered in a twisted version of ${\cal N} = 4$ supersymmetric Yang-Mills theory in four dimensions~\cite{Gukov-Witten}. They have also made an appearance in the context of the AdS/CFT correspondence between ${\cal N}=4$ SYM and type IIB supergravity~\cite{2}-\cite{8}, whereby the proposed action of the $SL(2,\mathbb Z)$ duality group on the parameters of a surface operator in~\cite{Gukov-Witten}, has been shown to be consistent in a dual type IIB supergravity description in~\cite{8}.

To date, there has not been an explicit way to prove that the parameters of a surface operator in the ${\cal N}=4$ gauge theory ought to transform as proposed in~\cite{Gukov-Witten}. Moreover, most examples involve only trivial embeddings of surface operators in spacetime; not much is known about the action of the $SL(2, \mathbb Z)$ duality group on the parameters of surface operators that are nontrivially-embedded.

Nonetheless, an explicit proof of the proposed transformation of parameters under an $SL(2,\mathbb Z)$ duality, was recently furnished in~\cite{SO abelian} for a pure, non-supersymmetric abelian gauge theory with trivially and nontrivially-embedded surface operators. In particular, it was found that for a nontrivially-embedded surface operator, exact $S$-duality (in the sense that the dual theory is of the same kind as the original theory, albeit with an inverted complexified gauge coupling) can only be maintained if its ``quantum'' parameter effectively vanishes, while the overall $SL(2,\mathbb Z)$ (or $\Gamma_0(2)$) duality holds up to a $c$-number at most, always. Also, the partition function and correlation functions of non-singular, gauge-invariant local operators transform, on curved four-manifolds, like modular forms of $SL(2,\mathbb Z)$ albeit with different modular weights.

\smallskip\noindent{\it A Summary of the Paper}

In this paper, we will generalise the analysis in~\cite{SO abelian} to superspace, and explicitly show that for an arbitrarily-embedded surface operator in a general, twisted ${\cal N}=2$ pure abelian gauge theory on any spin (or non-spin) four-manifold, the parameters transform naturally under the $SL(2,\mathbb Z)$ (or $\Gamma_0(2)$) duality of the theory. However, just as in the non-supersymmetric case, for a nontrivially-embedded surface operator, exact $S$-duality holds if and only if the ``quantum'' parameter effectively vanishes, while the overall $SL(2,\mathbb Z)$ (or $\Gamma_0(2)$) duality holds up to a $c$-number at most, regardless. Nevertheless, this observation sets the stage for forthcoming work~\cite{MC} that will provide, among other things, a physical proof of a remarkable mathematical result by Kronheimer and Mrowka~\cite{KM}, which, expresses a ``ramified'' analog\footnote{That is, an analog which includes surface operators that introduce a singularity in the $SU(2)$ or $SO(3)$ gauge field strength along a two-dimensional submanifold of the four-manifold.} of the Donaldson invariants of a four-manifold solely in terms of the ordinary Donaldson invariants. This mathematical result is also crucial to the formulation of an important structure theorem in Donaldson theory~\cite{KM,RR}, that first motivated the interpretation of the ordinary Donaldson invariants in terms of the ordinary Seiberg-Witten invariants in~\cite{monopoles}. We shall furnish a sketch of the general ideas behind this physical proof, and briefly explain how one can relate the ``ramified'' Donaldson invariants to the ordinary and ``ramified'' Seiberg-Witten invariants when we have a nontrivially and trivially-embedded surface operator, respectively. As a prelude to our forthcoming work, we will compute the effective interaction on the $u$-plane when the four-manifold is curved. We will also demonstrate a dependence on certain second Stiefel-Whitney classes, and the appearance of a $\textrm{Spin}^c$ structure in the associated low-energy Seiberg-Witten theory with surface operators, at points in moduli space where massless monopoles and dyons appear. In the process, we will stumble upon an interesting phase factor which one must include in the low-energy path-integral of the dual ``magnetic'' theory that is otherwise absent in the ``unramified'' case. As we shall see, the non-supersymmetric analysis in~\cite{SO abelian} will prove to be useful and insightful for our purposes.

\newsection{Surface Operators in ${\cal N}=2$ Abelian Gauge Theory}

\vspace{-0.3cm}

\newsubsection{Description of the Relevant Surface Operators}

\smallskip\noindent{\it The Parameters $\alpha$ and $\eta$}

In this paper, we shall consider surface operators that are supported on an arbitrary two-submanifold $D$ in a general, twisted ${\cal N}=2$ pure abelian gauge theory on some four-manifold $M$, where $D$ and $M$ are assumed to be closed and oriented. The surface operator is to be characterised by a gauge field solution that gives rise to a singular field strength as one approaches $D$. In addition, the gauge field solution must be invariant under rotations of the plane $D'$ normal to $D$.

An example of such a gauge field solution is
\be
{\cal A} = \alpha d\hat\theta,
\label{A}
\ee
where $\alpha$ is a ``classical'' parameter valued in $U(1)$,\footnote{Such a parameter of the gauge field ought to be valued in the (real) Lie algebra ${\frak u}(1)$. However, as explained in \cite{Gukov-Witten}, one can shift the parameter $\alpha \to \alpha + u$ in a particular gauge transformation, whereby $\text{exp}(2\pi u) =1$. The only invariant of such a gauge transformation is the monodromy $\text{exp}(-2\pi \alpha)$ of the gauge field $\cal A$ around a circle of constant $r$. Hence, $\alpha$ must take values in $\mathbb R / \mathbb Z$ instead.} and $\hat\theta$ is the angular component of the local coordinate $z = r e^{i\hat\theta}$ on $D'$ near $D$. Noting that $d(d\hat\theta) = 2 \pi \delta_D$ (where $\delta_D$ is a two-form delta function supported at the origin of $z$, that is also Poincar\'e dual to $D$), we find the corresponding field strength to be
\be
F = 2 \pi \alpha \delta_D.
\label{F}
\ee
As required, $F$ is singular as one approaches $D$. In such a case, we say that the theory is ``ramified''.

However, note from footnote~2 that we are free to shift $\alpha$ by $u$ via a gauge transformation. As such, this definition of $F$ appears to be unnatural. This can be remedied by lifting $\alpha$ in (\ref{F}) from $U(1)$ to ${\frak u}(1)$, such that it is no longer true that $\alpha \sim \alpha + u$, that is, $F$, when restricted to $D$, is ${\frak u}(1)$-valued. Equivalently, this corresponds to finding an extension of the $U(1)$-bundle $E$ on $M$ with connection $\cal A$, over $D$ (whereby due to the singularity along $D$, the bundle $E$ is originally defined on the complement of $D$ in $M$ only). Such an extension exists whenever $E$ is a $U(1)$-bundle on $M$. Thus, the definition of $F$ in (\ref{F}) actually does make sense.

Notice that since we have an extension of the bundle $E$ over $D$, we roughly have an abelian gauge theory in two dimensions on $D$. As such, one can introduce a two-dimensional theta-like angle $\eta$ as an additional ``quantum'' parameter, which enters in the Euclidean path-integral via the phase
\be
\textrm{exp}\left(i\eta \int_D F\right).
\label{eta term}
\ee
Notice that $\eta$ must therefore take values in $\mathbb R / \mathbb Z$, since the integrated first Chern class $\int_D F/2 \pi$ of the $U(1)$-bundle $E \to D$, is an integer. Just like $\alpha$, one can shift $\eta$ (by an integral lattice) whilst leaving the theory invariant.

\vspace{0.3cm}

\smallskip\noindent{\it A Point on Nontrivially-Embedded Surface Operators}

  More can also be said about the parameter $\alpha$ as follows. In the case when the surface operator is trivially-embedded in $M$, that is, $M = D' \times D$ and the normal bundle to $D$ is hence trivial, the self-intersection number
\be
D \cap D = \int_{M} \delta_D \wedge \delta_D
\label{DcapD}
\ee
vanishes. On the other hand, for a nontrivially-embedded surface operator supported on $D \subset M$, the normal bundle is nontrivial, and the intersection number is non-zero. The surface operator is then defined by the gauge field with singularity in (\ref{A}) in each normal plane.

When the intersection number is non-zero, or rather for nontrivially-embedded surface operators, there is a restriction on the values that $\alpha$ can take. To explain this, first note that since $F = 2 \pi \alpha \delta_D$ near $D$, we find, using (\ref{DcapD}), that $\int_D F/2 \pi  = \alpha \ D \cap D \ \textrm{mod}  \ \mathbb Z$. Since the integrated first Chern class $\int_D F/2 \pi$ is always an integer, we must have
\be
\alpha \ D \cap D \in \mathbb Z.
\label{intersection number}
\ee
This observation has an extension to the non-abelian case as follows~\cite{Gukov-Witten}: if $\alpha \to f(\alpha)$ is any real-valued linear function on $\frak t$ (the Lie algebra of the maximal torus of the non-abelian group) that takes integer values on the cocharacter lattice $\Lambda_{\textrm{cochar}}$, then
\be
f(\alpha) D \cap D \in \mathbb Z.
\label{intersection number 2}
\ee
In particular, the only gauge transformations that can be defined globally along $D$, are those that shift $\alpha$ in such a way as to maintain the condition (\ref{intersection number}) or (\ref{intersection number 2}). This point will be important later.

\smallskip\noindent{\it Supersymmetry and the Surface Operator}

Note that the surface operators defined above are actually supersymmetric and compatible with the ${\cal N} =2$ supersymmetry of the  abelian gauge theory to be discussed in this paper. In other words, their inclusion does not affect the supersymmetry of the underlying theory.  

To see this, first note that any supersymmetric field configuration of a theory must obey the conditions implied by setting the supersymmetric variations of the fermions to zero. In the original (untwisted) theory without surface operators, this implies that any supersymmetric field configuration must obey $F =0$ and $\nabla_\mu a = 0$, where $a$ is a scalar field in the ${\cal N} = 2$ vector multiplet~\cite{Marcos}. Let us assume for simplicity the trivial solution $a=0$ to the condition $\nabla_\mu a =0$ (so that the relevant moduli space is non-singular); this means that any supersymmetric field configuration must be consistent with $\it{irreducible}$ flat connections on $M$ that obey $F=0$. Consequently, the presence of any surface operator along $D$---which effects a monodromy  $\textrm{exp}({-2\pi \alpha})$ in the gauge field $\cal A$ as one traverses a loop that links the surface $D$---that is supposed to be supersymmetric and compatible with the underlying ${\cal N} =2$ supersymmetry,  ought to correspond to a $\it{flat}$ irreducible connection on the $U(1)$-bundle $E$ restricted to $M \setminus D$ which has the required singularity  along $D$;\footnote{This prescription of considering connections on the bundle $E$ restricted to $M \setminus D$ whenever one inserts a surface operator that introduces a field singularity along $D$,  is just a two-dimensional analog of the prescription one adopts when inserting an 't Hooft loop operator in the theory. See $\S$10.1 of~\cite{QFT2} for a detailed explanation of this.} equivalently, it must correspond to a $\it{flat}$ irreducible connection on a $U(1)$-bundle $E'$ over $M$ whose field strength is $F' = F - 2\pi \alpha \delta_D$, where $F = d {\cal A}$ is the field strength of the bundle $E$ over  $M$ that is hence singular along $D$.\footnote{To justify this statement, note that the instanton number $\tilde k$ of the bundle $L$ over $X \setminus D$ is (in the mathematical convention)  given by $\tilde k = k +  \alpha l - (\alpha^2 / 2) D \cap D$, where $k$ is the instanton number of the bundle $L$ over $X$ with curvature $F$, and $l = \int_D F / 2\pi$ is the monopole number  ($\it cf$. eqn.~(1.7) of~\cite{Mrowka} for a $U(1)$-bundle). On the other hand,  the instanton number $k'$ of the bundle $L'$ over $X$ with curvature $F' = F - 2\pi \alpha \delta_D$ is (in the physical convention) given by $k' = - {1\over 8 \pi^2} \int_X F' \wedge F' = k +  \alpha l - (\alpha^2 / 2) D \cap D$. Hence, we find that the expressions for $\tilde k$ and $k'$ coincide, reinforcing the notion that the bundle $L$ over $X \setminus D$ can be equivalently interpreted as the bundle $L'$ over $X$. Of course, for $F'$ to qualify as a nontrivial field strength, $D$ must be a homology cycle of $X$, so that $\delta_D$ (like $F$) is in an appropriate cohomology class  of $X$.}  In other words, a supersymmetric surface operator will correspond to a gauge field solution $\cal A$ that satisfies $F' =0$; that is, $F = 2 \pi \alpha \delta_D$.  Indeed, this is the condition (\ref{F}) that characterises our surface operator in the first place. 

As a result of the singularity (\ref{F}) when one includes a surface operator in the theory, the effective field strength in the Lagrangian that contributes non-vanishingly to the path-integral must be a shifted version of the field strength $F$. (We shall explain this in greater detail below). At any rate, note that since the surface operator does not introduce any singularities in the other fields of the underlying theory, it suffices to modify only the field strength to obtain the effective Lagrangian. That being said, in a different theory whereby supersymmetric configurations involve not just the field strength $F$ but also the other fields, a supersymmetric surface operator would give rise to a singularity along $D$ in the other fields as well. For example, in the pure ${\cal N}=4$  theory considered in~\cite{Gukov-Witten}, supersymmetric configurations involve the Higgs field $\phi$ in addition to the field strength $F$. Consequently, the inclusion of a surface operator in the ${\cal N} =4$ theory that is supposedly supersymmetric, will also give rise to a singularity in $\phi$ along $D$.

\newsubsection{Action of Duality on Trivially-Embedded Surface Operators}

\smallskip\noindent{\it Action of $S$-duality}

We shall now discuss the case of a trivially-embedded surface operator---with $\it a$-$\it{priori}$ non-vanishing parameters $(\alpha, \eta)$---in a general, twisted ${\cal N}=2$ pure abelian gauge theory on $\it{any}$ smooth four-manifold $M$. Our first objective is to prove explicitly that the parameters transform as
\be
(\alpha, \eta) \to (\eta, - \alpha)
\label{parameter transform 1}
\ee
under the $S$-duality transformation $S : \tau (a) \to - 1/ \tau (a)$ of the supersymmetric gauge theory. Here, $\tau (a) = {\Theta (a) / 2\pi}  + {4\pi i / g^2(a)}$ is the effective complexified gauge coupling in the vacuum parameterised by $a$. To this end, we shall adapt the approach of~\cite{SO abelian} to superspace.

Firstly, let us note that the most general action of an ${\cal N}=2$ pure abelian gauge theory in Minkowski space can be written as~\cite{Seiberg}
\be
I = -{1\over 2\pi} \textrm{Im} \left[ \int d^4 x  d^2 \theta d^2 {\bar\theta} \ {\partial {\cal F} \over {\partial A}}  \bar A + \int d^4 x  d^2 \theta \ {1\over 2} {\partial^2 {\cal F} \over{\partial A^2}} \ W^{\alpha}W_{\alpha} \right],
\ee
where $A$ and $W$  are a chiral superfield and a chiral spinorial (abelian) superfield strength in ${\cal N}=1$ superspace, whose components together make up the ${\cal N} =2$ vector multiplet that defines the pure ${\cal N}=2$ abelian gauge theory. Also, $\cal F$ is a holomorphic function of $A$, from which one can obtain the K\"ahler potential of the one-complex dimensional moduli space of the theory as the scalar component of
\be
K = \textrm{Im} \left( {\partial{\cal F}\over {\partial A}} \bar A \right).
\ee
The metric on the moduli space can thus be written as
\be
(ds)^2 = g_{a\bar a} da d \bar a =  \textrm{Im} \tau(a) da d \bar a,
\ee
where $a$ is the scalar component of $A$, and $\tau(a)$---which is a holomorphic function of $a$---is given by
\be
\tau(a) = {\partial^2 {\cal F}(a) \over {\partial a^2}}.
\ee
Consequently, one can rewrite the most general action as
\be
I = -{1\over 2\pi} \textrm{Im} \left[ \int d^4 x  d^2 \theta d^2 {\bar\theta} \ {\partial {\cal F} \over {\partial A}}  \bar A\right] - {1\over 4\pi}\textrm{Im}\left[\int d^4 x  d^2 \theta \ \tau(A) \ W^{\alpha}W_{\alpha} \right],
\label{general action}
\ee
where
\be
\tau(A) = {\partial^2 {\cal F}(A)\over \partial A^2}.
\ee

Now, let us for ease of illustration, consider the case of ${\cal F}(A) = {1\over 2}\tau_{cl} A^2$, where $\tau_{cl} = {\Theta / 2\pi}  + {4\pi i / g^2}$ is the classical complexified gauge coupling that is independent of $a$ and is thus constant. In this case, one has $\tau(A) = \tau_{cl}$, and the $\it{second}$ term in (\ref{general action}) can be written (after eliminating an auxiliary field $D$ using its equation of motion) as
\be
I_2 =  {1\over g^2}\int d^4x \ \left({1\over 2} F_{\mu \nu} F^{\mu\nu} + 2 i \lambda \sigma^\mu \partial_\mu \bar\lambda \right)  - {\Theta \over 8 \pi^2}\int_M F \wedge F,
\ee
where we have used the fact that $W$ has an expansion
\be
W_\alpha = -i \lambda_\alpha (x) + \theta_\alpha D(x) + i (\sigma^{\mu \nu} \theta)_\alpha F_{\mu\nu}(x) + \theta\theta (\sigma^\mu \partial_\mu \bar \lambda(x))_\alpha,
\label{W-expansion}
\ee
and the identity
\be
(\sigma^{\mu\nu})^{\alpha\beta} (\sigma^{\rho\sigma})_{\alpha\beta} = {1\over 2} (g^{\mu \rho}g^{\nu \sigma} - g^{\mu \sigma}g^{\nu \rho}) - {i\over 2} \epsilon^{\mu\nu\rho\sigma}.
\label{matrix identity}
\ee
In the above, the $\sigma^\mu$'s are the Pauli matrices such that ${\sigma^{\mu \nu}}_{\alpha}{}{}^\beta = {1\over 4} (\sigma^\mu_{\alpha \dot\alpha} {\bar\sigma}^{\nu \dot\alpha \beta} - {\sigma}^\nu_{\alpha \dot\alpha}{\bar\sigma}^{\mu \dot\alpha \beta})$, where $\alpha, \dot\alpha = 1, 2$. Also, our convention is such that $\epsilon^{0123} =1$.

The action $I_2$ is equivalent to a dual action $I_{2,D}$ in the dual fields $F_D$, $\lambda_D$ and ${\bar\lambda}_D$   with gauge coupling $\tau_D = -{1/ \tau}$, as first shown in~\cite{Seiberg-Witten}. Moreover, since the ordinary instanton number $-{1\over 8 \pi^2}\int_M F \wedge F$ is always an integer, the theory is also invariant under $\Theta \to \Theta + 2\pi$, that is,  $\tau \to \tau +1$. Our immediate objective is to show that a similar duality holds when we include surface operators in the theory with action $I_2$.

Note at this point that when we include a surface operator along $D \subset M$, a singularity in the field strength of the form $2 \pi \alpha \delta_D$ will be introduced near D. In addition, one must also include in the Euclidean action the topological term $-i \eta \int_D F$, as mentioned earlier. However, notice that since the kinetic term of the gauge field in $I_2$ has a positive-definite real part, the Euclidean path-integral (which is what we would eventually be interested in) would be non-vanishing only if the contributions to the kinetic term  are non-singular. As such, one can equivalently study the action with field strength $F' = F - 2\pi \alpha \delta_D$ instead of $F$, whenever we have a surface operator along $D$. This important fact was first pointed out in~\cite{Gukov-Witten}, and further exploited in~\cite{SO abelian} to prove the $S$-duality in the non-supersymmetric case. Hence, in the presence of a surface operator along $D$, the effective Minkowskian action to consider takes the form
\be
I'_2 =  {1\over g^2}\int d^4x \ \left({1\over 2} F'_{\mu \nu} {F'}^{\mu\nu} + 2 i \lambda \sigma^\mu \partial_\mu \bar\lambda \right)  - {\Theta \over 8 \pi^2}\int_M F' \wedge F' + \eta\int_D F'.
\label{2.16}
\ee

In turn, note that $I'_2$ can be written in superspace as
\be
I'_2 = -{1\over 4\pi}\textrm{Im}\left[\int d^4 x  d^2 \theta \ \tau(A) \ {W'}^{\alpha}W'_{\alpha} \right] + {1 \over 2 \pi} \textrm{Im} \left [\int d^4 x d^2 \theta \ [i 2 \pi \eta (\delta_D)_{\mu \nu} (\sigma^{\mu\nu}\theta)]^\alpha W_\alpha \right],
\label{I'_2}
\ee
where $W'_\alpha = W_\alpha - i 2 \pi \alpha (\delta_D)_{\mu\nu} (\sigma^{\mu\nu}\theta)_\alpha$, that is,
\be
W'_\alpha = -i \lambda_\alpha (x) + \theta_\alpha D(x) + i (\sigma^{\mu \nu} \theta)_\alpha (F - 2\pi \alpha \delta_D)(x)_{\mu\nu} + \theta\theta (\sigma^\mu \partial_\mu \bar \lambda(x))_\alpha.
\label{W'-expansion}
\ee
Note that in order to arrive at (\ref{I'_2}), we have used the identities $\int d^2 \theta \ i(\delta_D)_{\mu\nu} (\sigma^{\mu \nu}\theta)^\alpha  \ i(\sigma^{\rho \sigma}\theta)_\alpha F_{\rho\sigma} \newline = -{1\over 2} (\sigma^{\mu\nu})^{\alpha\beta}(\sigma^{\rho\sigma})_{\alpha\beta} (\delta_D)_{\mu\nu} F_{\rho \sigma}$ and (\ref{matrix identity}), the relation $\eta\int_D F' = \int_M \delta_D \wedge F'$, and the fact that for trivially-embedded surface operators, the term $-2\pi \eta \alpha \int_M \delta_D \wedge \delta_D$ that would generically appear, can be set to zero, since $D\cap D = 0$ in this case.

Notice that since $W'$ persists as a (spinorial) chiral superfield,\footnote{This is true because one can also write $W'_\alpha = -{1\over 4} \bar D \bar D D_\alpha V'$ for some vector superfield $V'$, where ${D}_{\alpha} =  {\partial \over \partial {\theta}^{\alpha}} + i  \sigma^\mu_{\alpha {\dot\alpha}}{\bar \theta}^{\dot \alpha} \partial_\mu$ and $\bar D_{\dot\alpha} = -{\partial \over \partial {\bar \theta}^{\dot\alpha}} - i  {\theta}^{\alpha}\sigma^\mu_{\alpha {\dot\alpha}}\partial_\mu$, such that $\bar D_{\dot\beta} W' = 0$; that is, $W'$ is also a chiral superfield.} the first term in $I'_2$ remains invariant under   supersymmetry transformations. Similarly, since the second term just corresponds to the topological term $ \eta \int_M \delta_D \wedge F'$, it will be invariant under supersymmetry transformations as well. Therefore, $I'_2$ persists as a valid ${\cal N} =2$ supersymmetric action.

Another point to note at this juncture is that apart from satisfying the chirality condition ${\bar D}_{\dot\beta} W_\alpha = 0$, $W$ also satisfies the superspace Bianchi identity $\textrm{Im}({D}_{\alpha} W^\alpha) = 0$. In order to implement this condition on $W$, we can add to the action (\ref{I'_2}) the Lagrange multiplier term
\be
I_m  =  -{1\over 8 \pi} \textrm{Im} \left[ \int d^4x d^2 \theta d^2\bar\theta \ V_D  D_\alpha W^\alpha \right]
 = -{1\over 2 \pi} \textrm{Im} \left[ \int d^4x d^2\theta \ (W_D)^{\alpha} W_\alpha  \right],
\ee
where $V_D$ is a real Lagrange multiplier superfield with chiral  field strength $(W_D)_{\alpha} = - {1\over 4} {\bar D}^2 D_\alpha V_D$, such that one can write
\be
(W_D)_\alpha = -i (\lambda_D)_\alpha (x) + \theta_\alpha {\tilde D}(x) + i (\sigma^{\mu \nu} \theta)_\alpha (F_D)_{\mu\nu}(x) + \theta\theta [\sigma^\mu \partial_\mu (\bar \lambda_D)(x)]_\alpha,
\label{W_D-expansion}
\ee
where the subscript ``D'' indicates that the above fields are dual to the corresponding fields in $W$ of (\ref{W-expansion}). (This statement will be justified shortly).

Consequently, the action can now be written as
\be
{\cal I} = I'_2 + I_m = -{1\over 4\pi}\textrm{Im}\left[\int d^4 x  d^2 \theta \ \tau(A) \ {W'}^{\alpha}W'_{\alpha} \right] - {1\over 2 \pi} \textrm{Im} \left[ \int d^2\theta \ (W'_D)^{\alpha} W_\alpha  \right],
\ee
where $(W'_D)_\alpha = (W_D)_\alpha -  i 2 \pi \eta (\delta_D)_{\mu\nu} (\sigma^{\mu\nu}\theta)_\alpha$, that is,
\be
(W'_D)_\alpha = -i (\lambda_D)_\alpha (x) + \theta_\alpha {\tilde D}(x) + i (\sigma^{\mu \nu} \theta)_\alpha (F_D - 2\pi \eta \delta_D)(x)_{\mu\nu} + \theta\theta [\sigma^\mu \partial_\mu (\bar \lambda_D)(x)]_\alpha.
\label{W_D'-expansion}
\ee

Let us now define the following chiral (spinorial) superfield
\be
{\widetilde W}_\alpha = W'_\alpha + {1\over \tau(A)} (W'_D)_\alpha,
\ee
where $W'_{\alpha}$ and $(W'_D)_\alpha$ are given in (\ref{W'-expansion}) and (\ref{W_D'-expansion}), respectively. After a straightforward computation, we find that
\begin{eqnarray}
{\cal I} &= & -{1\over 4 \pi} \textrm{Im} \left[ \int d^4 x d^2 \theta \ \tau(A) {\widetilde W}^{\alpha} {\widetilde W}_\alpha \right] - {1\over 4 \pi} \textrm{Im} \left[ \int d^4 x d^2\theta \ \left( -{1\over \tau(A)}\right) (W'_D)^{\alpha} (W'_D)_\alpha \right] \nonumber \\
 & & \quad - {1\over 2 \pi} \textrm{Im} \left[ \int d^4x d^2\theta \ [i 2 \pi \alpha (\delta_D)_{\mu \nu} (\sigma^{\mu\nu}\theta)]^\alpha (W'_D)_\alpha \right].
\end{eqnarray}
By integrating ${\widetilde W}$ out in superspace, we can simplify $\cal I$ to
\be
{\cal I} =  -{1\over 4 \pi} \textrm{Im} \left[ \int d^4 x d^2\theta \ \left( -{1 \over \tau(A)}\right) (W'_D)^{\alpha} (W'_D)_\alpha \right] - {1\over 2 \pi} \textrm{Im} \left[ \int d^4x d^2\theta \ [i 2 \pi \alpha (\delta_D)_{\mu \nu}
(\sigma^{\mu\nu}\theta)]^\alpha (W_D)_\alpha \right],
\label{cal I}
\ee
where we have again made use of the fact that the term $2\pi \eta \alpha \int_M \delta_D \wedge \delta_D$ that would generically appear can be set to zero, since $D\cap D = 0$ for a trivially-embedded surface operator.

By comparing (\ref{cal I}) with (\ref{I'_2}), we find that the transformations $\tau(A) \to - {1 / \tau(A)}$, $W \to W_D$, $\alpha \to \eta$ and $\eta \to -\alpha$, map $I'_2$ to $\cal I$. Since $\cal I$ is supposed to be physically equivalent to $I'_2$, these transformations represent a duality of the theory with action $I'_2$. It is in this sense that the field components of $W_D$ are dual to those of $W$, as mentioned earlier.

Notice that our above derivation of the duality does not make any reference to the explicit form of $\tau(A)$. As such, one can generalise the computation to $\it{any}$ $\tau(A)$ beyond $\tau(A) = \tau_{cl}$, and still arrive at the same conclusion. In order to obtain the corresponding spacetime expression of $I'_2$, one would just need to perform the $\theta$-integration of the second term in (\ref{general action}), and again replace $F$ everywhere with $F'= F - 2 \pi \alpha \delta_D$. We shall do that shortly, but let us proceed to discuss the first term in (\ref{general action}) now.

To this end, first note that the spacetime contributions of the first term in (\ref{general action}) do not contain the field strength $F$. Therefore, by including a surface operator in the theory---unlike what had to be done with $W$ of the second term---there is no need to shift the integrand $\partial {\cal F}/ \partial A$ or $\bar A$. Thus, the analysis follows as in the usual case without surface operators in~\cite{Seiberg-Witten}. Nonetheless, we shall, for self-containment of the paper, review the computation anyhow.

Now, if we define $A_D$ to be the chiral superfield dual to $A$ such that
\be
A_D = h(A) = {\partial {\cal F}(A) \over \partial A}
\ee
for some function $h(A)$ that is holomorphic in $A$, then one can write the first term in (\ref{general action}) as
\be
-{1\over 2\pi} \textrm{Im} \left[ \int d^4 x  d^2 \theta d^2 {\bar\theta} \ h(A) \bar A\right] = -{1\over 2\pi} \textrm{Im} \left[ \int d^4 x  d^2 \theta d^2 {\bar\theta} \ h_D(A_D) {\bar A}_D\right]
\label{equal}
\ee
if $h_D(A_D) = h_D(h(A)) = -A$. Consequently, if one also defines
\be
h_D(A_D) = {\partial {\cal F}(A_D) \over \partial A_D} = -A,
\ee
we find (from $\tau(A) = {\partial^2 {\cal F}(A) / \partial A^2}$) that
\be
- {1 \over \tau(A)} = - {1 \over {\partial h(A) \over \partial A}} = {\partial h_D(A_D) \over \partial A_D} = {\partial^2 {\cal F}(A_D) \over \partial A_D^2}=  \tau_D(A_D).
\ee

Altogether, this means that the theory with general action
\begin{eqnarray}
I' & = &  -{1\over 2\pi} \textrm{Im} \left[ \int d^4 x  d^2 \theta d^2 {\bar\theta} \ h(A) \bar A \right] - {1\over 4\pi}\textrm{Im} \left[\int d^4 x  d^2 \theta \ \tau(A)  {W'}^{\alpha}{W'}_{\alpha} \right] \nonumber \\
&& \quad + {1 \over 2 \pi} \textrm{Im} \left [\int d^4 x d^2 \theta \ [i 2 \pi \eta (\delta_D)_{\mu \nu} (\sigma^{\mu\nu}\theta)]^\alpha W_\alpha \right],
\label{I'}
\end{eqnarray}
is physically $\it{equivalent}$ to the theory with general action
\begin{eqnarray}
I'_D & = &  -{1\over 2\pi} \textrm{Im} \left[ \int d^4 x  d^2 \theta d^2 {\bar\theta} \ h_D(A_D) {\bar A}_D \right] - {1\over 4\pi}\textrm{Im} \left[\int d^4 x  d^2 \theta \ \tau_D(A_D)  (W'_D)^{\alpha}(W'_D)_{\alpha} \right] \nonumber \\
&& \quad - {1 \over 2 \pi} \textrm{Im} \left [\int d^4 x d^2 \theta \ [i 2 \pi \alpha (\delta_D)_{\mu \nu} (\sigma^{\mu\nu}\theta)]^\alpha (W_D)_\alpha \right].
\label{I'_D}
\end{eqnarray}
In other words, the transformations $A \to A_D$, $h(A) \to h_D (A_D)$, $\tau(A) \to \tau_D(A_D)$, $W \to W_D$, $\alpha \to \eta$ and $\eta \to -\alpha$, are duality transformations of the general, ${\cal N} = 2$ pure abelian gauge theory with a trivially-embedded surface operator in Minkowski space.

By expanding in component fields and performing the $\theta$-integrations in (\ref{I'}) and (\ref{I'_D}), we can write the Minkowskian Lagrangian densities $L'$ and $L'_D$ of the actions $I'$ and $I'_D$, respectively, as
\begin{eqnarray}
\label{L'}
L' & = & {1 \over g^2} {F'}\wedge \star {F'} - {\Theta \over 8 \pi^2} (F' \wedge F') + {1\over 2\pi} \left[(\textrm{Im} \tau) \partial_\mu a \partial^\mu \bar a + i (\textrm{Im}\tau)  \lambda^m \sigma^\mu \partial_\mu {\bar \lambda}_m\right]  \nonumber \\
&& + {\sqrt 2 \over 8 \pi} \textrm{Im}\left[ {d\tau \over da} \lambda^m \sigma^{\mu\nu} \lambda_m F'_{\mu\nu} \right] + {\sqrt 2 \over 8 \pi} \textrm{Im} \left[ {d\tau \over da} \lambda^m \lambda^n D_{m n} \right] - {1 \over 8 \pi} \textrm{Im} \tau D_{m n}D^{m n} \\
&&  - {1 \over 24 \pi} \textrm{Im} \left[{d^2 \tau \over da^2} (\lambda^m \lambda^n)(\lambda_m \lambda_n) \right] + \eta (\delta_D \wedge F') \nonumber
\end{eqnarray}
and
\begin{eqnarray}
L'_D & = & {1 \over g^2_D} {F'_D}\wedge \star {F'_D} - {\Theta_D \over 8 \pi^2} (F'_D \wedge F'_D)  + {1\over 2\pi} \left[(\textrm{Im} \tau_D) \partial_\mu a_D \partial^\mu \bar a_D + i (\textrm{Im}\tau_D)  \lambda_D^m \sigma^\mu \partial_\mu {\bar\lambda}_{D m}\right]  \nonumber \\
&& + {\sqrt 2 \over 8 \pi} \textrm{Im}\left[ {d\tau_D \over da_D} \lambda_D^m \sigma^{\mu\nu} \lambda_{D m} {F'_D}_{\mu\nu} \right] + {\sqrt 2 \over 8 \pi} \textrm{Im} \left[ {d\tau_D \over da_D} \lambda_D^m \lambda_D^n {\tilde D}_{m n} \right] - {1 \over 8 \pi} \textrm{Im} \tau_D {\tilde D}_{m n}{\tilde D}^{m n} \nonumber \\ &&  - {1 \over 24 \pi} \textrm{Im} \left[{d^2 \tau_D \over da_D^2} (\lambda_D^m \lambda_D^n)(\lambda_{D m} \lambda_{D n}) \right] - \alpha (\delta_D \wedge F'_D),
\label{Dual L}
\end{eqnarray}
where $F' = F - 2\pi \alpha \delta_D$, $F'_D = F_D - 2\pi \eta \delta_D$ and $\tau_D(a_D) = {\Theta_D(a_D) / 2\pi} + {4 \pi i / g^2_D(a_D)} = -1/\tau(a)$. (Here, $a_D$ is the scalar component of $A_D$). Also, the indices ``m'' and ``n'' are associated to the internal $SU(2)$ global $R$-symmetry of the theory, and $D$ is again an auxiliary field.

We would now like to twist the theory to make it topological so that it can be defined on an arbitrary four-manifold. To this end, let us first wick-rotate $I'$ and $I'_D$ into Euclidean actions. This in particular, would introduce a factor of $-i$ in the $\Theta$, $\alpha$ and $\eta$ terms in $L'$ and $L'_D$. Since the total, global symmetry group of the theory is ${\cal H} = SU(2)_+ \times SU(2)_- \times SU(2)_R \times U(1)_R$---where ${\cal K} = SU(2)_+ \times SU(2)_-$ is the rotation group in four-dimensional Euclidean space and $U(2)_R \simeq SU(2)_R \times U(1)_R$ is the internal symmetry group---the standard twisting recipe~\cite{TQFT} would then entail a redefinition of the rotation group to  ${\cal K}' = SU'(2)_+ \times SU(2)_-$, where $SU'(2)_+$ is the diagonal combination of $SU(2)_+ \times SU(2)_R$. The spins of the fields with respect to the new rotation group ${\cal K}'$ would then be such that two of the original eight supercharges of the ${\cal N}=2$ supersymmetry would now transform as zero-forms (that is, scalars), $\lambda$ would now transform as a one-form $\psi$, and $\bar \lambda$ would decompose into a linear combination of a  zero-form $\phi$ and a self-dual two-form $\chi$. The nilpotent topological supercharge---which one can now define on $\it{any}$ four-manifold---is a linear combination of the two supercharges that transform as zero-forms. Note that what we have actually done in the twisting procedure is to couple the background fields to the $SU(2)_R$ global symmetry current of the theory. Since this current is invariant under the transformations that map $I'$ to $I'_D$, it means that the actions ${\widehat I}'$ and ${\widehat I}'_D$ will also be physically equivalent, where ${\widehat I}'$ and ${\widehat I}'_D$ are the twisted variants of the (Euclidean version) of the actions $I'$ and $I'_D$, respectively. Moreover, ${\widehat I}'$ can be related to ${\widehat I}'_D$ by the same duality transformations that map $I'$ to $I'_D$.

Since we will not need to refer to the explicit form of ${\widehat I}'$ or ${\widehat I}'_D$, or the corresponding supersymmetric variations of fields which leave them invariant, we shall, for brevity, not state them here  in this paper. Nonetheless, we can conclude that under the $S$-duality transformation $\tau(a) \to - 1 / \tau(a)$ of the general, twisted ${\cal N}=2$ pure abelian gauge theory with a trivially-embedded surface operator on an arbitrary four-manifold $M$, the parameters will transform as $(\alpha, \eta) \to (\eta, - \alpha)$, as claimed.

\vspace{0.3cm}

\smallskip\noindent{\it Action Under a Shift in Theta-Angle}

Our second objective is to prove that the parameters $(\alpha, \eta)$ transform under the symmetry $T: \tau(a) \to \tau(a) + 1$ as
\be
(\alpha, \eta) \to (\alpha, \eta - \alpha)
\label{shift}
\ee
for $M$ a $\it{spin}$ manifold.

As mentioned above, the theta-angle term of the topological action ${\widehat I}'$ will be given by
\be
{\widehat I}'_{\Theta} = {i \Theta \over {8 \pi^2}} \int_M  F' \wedge F'.
\label{theta-term}
\ee
This can also be written as
\be
{\widehat I}'_{\Theta}  =  i \Theta {\bf N},
\ee
where
\be
{\bf N} = {1 \over 2} c_1({\cal L})^2  - \alpha \frak m,
\label{N}
\ee
and $\cal L$ is the $U(1)$-bundle whose curvature is given by $F$, while $\frak m = \int_D (F/ 2\pi)$ is the ``magnetic charge'' associated with the flux through $D$. A term $(\alpha^2 / 2) D \cap D$ that generically appears in $\bf N$ has been set to zero above, since we are considering surface operators which are trivially-embedded at this point. Also, the first term $(1/2) c_1({\cal L})^2$ is always an integer, since $M$ is defined to be spin.

Next, consider the term
\be
{\widehat I}'_{\eta}  = -i \eta \int_D F' = -i \eta \int_M \delta_D \wedge F',
\ee
which is the only term in the total action that can potentially cancel the variation of ${\widehat I}'_{\Theta}$ under the transformation $T: \Theta \to \Theta + 2 \pi$. It can also be written as
\be
{\widehat I}'_{\eta}  =  -2 \pi i \eta {\frak m},
\label{I_shift}
\ee
where a term $ 2 \pi i \alpha \eta D \cap D$ has been set to zero in ${\widehat I}'_{\eta}$ above, since we are considering only trivially-embedded surface operators here.

Thus, the sum of the two contributions to the total action is then
\be
{\widehat I}'_{\Theta} + {\widehat I}'_{\eta} =  i \Theta {\bf N} - 2 \pi i \eta {\frak m}.
\ee
The variation in ${\widehat I}'_{\Theta}$ under $T: \Theta \to \Theta + 2 \pi$ is (mod $2 \pi i \mathbb Z$)
\be
\Delta {\widehat I}'_{\Theta} =  - 2 \pi i \alpha \frak m.
\ee
Hence, in order for the total contribution ${\widehat I}'_{\Theta} + {\widehat I}'_{\eta}$ to be invariant, one must have the transformation $(\alpha, \eta) \to (\alpha, \eta -\alpha)$ under $T:\tau(a) \to \tau(a) +1$, as claimed.

If $M$ is $\it{not}$ a spin manifold, the original theory without surface operators is only invariant under $T^2: \tau(a) \to \tau(a) +2$. This is because $c_1({\cal L})^2$ is no longer an even integer. Repeating the above analysis, we find that the parameters must transform as
\be
(\alpha, \eta) \to (\alpha, \eta - 2\alpha)
\label{shift non-spin}
\ee
under $T^2:\tau(a) \to \tau(a) +2$, when $M$ is non-spin.

\vspace{0.3cm}

\smallskip\noindent{\it Action Under Overall Duality}

Note that the $SL(2,\mathbb Z)$ duality group is an infinite discrete group which acts on $\tau = \tau(a)$ as
\be
\tau \to {{(a \tau + b)} \over {(c \tau +d)}},  \qquad \left(\begin{array}{ccc} a & & b \\ c & & d \end{array} \right) \in SL(2,\mathbb Z).
\ee
It is generated by the transformations $S: \tau \to -1 /\tau$ and $T: \tau \to \tau +1$, where
\be
S = \left(\begin{array}{ccc} 0 & & 1 \\ -1 & & 0 \end{array} \right), \qquad T = \left(\begin{array}{ccc} 1 & & 1 \\ 0 & & 1 \end{array} \right).
\ee
From (\ref{parameter transform 1}) and (\ref{shift}), we find that $\alpha$ and $\eta$ transform as
\be
(\alpha, \eta) \to (\alpha, \eta) {\frak M}^{-1},
\ee
where $\frak M$ is $S$ or $T$, accordingly. Therefore, this is true for any ${\frak M} \in SL(2,\mathbb Z)$. Hence, we see that $(\alpha, \eta)$ transform naturally under the $SL(2,\mathbb Z)$ duality of the general, twisted ${\cal N}=2$ pure abelian gauge theory on a spin manifold $M$. In particular, $(\alpha, \eta)$ transform under $S$-duality just like magnetic and electric charges do, respectively.

On the other hand, consider the congruence subgroup $\Gamma_0(2)$ that is generated by the transformations $S$ and $ST^2S$, that is,
\be
S = \left(\begin{array}{ccc} 0 & & 1 \\ -1 & & 0 \end{array} \right), \qquad ST^2S = \left(\begin{array}{ccc} -1 & & 0 \\ 2 & & -1 \end{array} \right).
\ee
From (\ref{parameter transform 1}) and (\ref{shift non-spin}), we find that $\alpha$ and $\eta$ transform as
\be
(\alpha, \eta) \to (\alpha, \eta) {\frak M}^{'-1},
\ee
where ${\frak M}'$ is $S$ or $ST^2S$, accordingly. Therefore, this is true for any ${\frak M}' \in \Gamma_0(2)$. Hence, we see that $(\alpha, \eta)$ transform naturally under a $\Gamma_0(2)$ duality of the general, twisted ${\cal N}=2$ pure abelian gauge theory on a $\it{non}$-$\it{spin}$ manifold $M$. Nonetheless, $(\alpha, \eta)$ continue to transform under $S$-duality just like magnetic and electric charges do, respectively.

\vspace{0.3cm}

\smallskip\noindent{\it Analogy with $A$ and $A_D$}

As shown above, the theory is invariant under the transformations $A \to A_D$ and $h(A) \to h_D(A)$. Since $h(A) = A_D$ and $h_D(A_D) = -A$, it would mean that the following is a duality transformation of the theory:
\be
\left(\begin{array}{c} A \\ A_D \end{array} \right) \to  \left(\begin{array}{ccc} 0 & & 1 \\ -1 & & 0 \end{array} \right)\left(\begin{array}{c} A \\ A_D \end{array} \right).
\label{2.48}
\ee

In addition, let us now shift $h(A) \to h(A) + m A$, where $m$ is a real constant. Then, the first term in (\ref{I'}) will shift by $m (\textrm{Im}[A \bar A]_{D})/2\pi$. Since the $D$-term $[A \bar A]_{D}$ of $A \bar A$ is real, this shift vanishes. On the other hand, since $\tau(A) = {\partial h(A) / \partial A}$, the presence of the second term in (\ref{I'}) will shift (the Euclidean version of) $L'$ by
\be
L' \to L' + {i m \over 4 \pi} \int_M F'\wedge F'.
\ee
Comparing this with ${\widehat I}'_{\Theta}$ in (\ref{theta-term}), we see that the above transformation shifts $\Theta$ by $2\pi m$. Hence, if $m =1$ (or $2$), the theory is invariant, since the transformation $\Theta \to \Theta + 2\pi$ (or $\Theta \to \Theta + 4\pi$) is a symmetry of the theory for $M$ spin (or non-spin), as explained earlier. In other words, the transformation
\be
\left(\begin{array}{c} A \\ A_D \end{array} \right) \to  \left(\begin{array}{ccc} 1 & & 0 \\ 1 & & 1 \end{array} \right)\left(\begin{array}{c} A \\ A_D \end{array} \right)
\label{2.50}
\ee
is a duality transformation of the theory when $M$ is spin, while the transformation
\be
\left(\begin{array}{c} A \\ A_D \end{array} \right) \to  \left(\begin{array}{ccc} 1 & & 0 \\ 2 & & 1 \end{array} \right)\left(\begin{array}{c} A \\ A_D \end{array} \right)
\label{2.51}
\ee
is a duality transformation of the theory when $M$ is non-spin.

Since the $2 \times 2$ matrices appearing in (\ref{2.48}) and (\ref{2.50}) together generate the entire $SL(2,\mathbb Z)$ group, we conclude that for $M$ spin, $A$ and $A_D$ transform as
\be
(A_D, A) \to (A_D, A) {\frak M}^{-1},
\ee
for any ${\frak M} \in SL(2,\mathbb Z)$. Hence, we see that $(A_D, A)$ transform naturally under the $SL(2,\mathbb Z)$ duality of the general, twisted ${\cal N}=2$ pure abelian gauge theory on a spin manifold $M$.

On the other hand, since the $2 \times 2$ matrices appearing in (\ref{2.48}) and (\ref{2.51}) together generate the entire $\Gamma_0(2)$ group, we conclude that for $M$ non-spin, $A$ and $A_D$ transform as
\be
(A_D, A) \to (A_D, A) {{\frak M}'}^{-1},
\ee
for any ${\frak M}' \in \Gamma_0(2)$. Hence, we see that $(A_D, A)$ transform naturally under the $\Gamma_0(2)$ duality of the general, twisted ${\cal N}=2$ pure abelian gauge theory on a $\it{non}$-$\it{spin}$ manifold $M$.

Notice that $(\alpha, \eta)$ transform similarly to $(A_D, A)$ under the $SL(2,\mathbb Z)$ or $\Gamma_0(2)$ duality of the theory on spin or non-spin four-manifolds. It is in this sense that the actions $I'$ and $I'_D$---expressed in the superfields $A$ and $A_D$---are said to define the ``electric'' and ``magnetic'' frames of the underlying theory, respectively.

\newsubsection{Action of Duality on Nontrivially-Embedded Surface Operators}

\smallskip\noindent{\it Action Under $S$-duality}

The analysis for the case of a nontrivially-embedded surface operator is similar to the one above except for one crucial difference; in the presence of a nontrivially-embedded surface operator, we have instead the following physical equivalence of actions
\be
{\widehat I}'_D \equiv {\widehat I}' - 2 \pi i \eta \alpha \int_M \delta_D \wedge \delta_D.
\label{crucial diff}
\ee
Note that the term $2 \pi i \eta \alpha \int_M \delta_D \wedge \delta_D$ is non-vanishing in a generic situation because $D\cap D \neq 0$. However, for $\it{exact}$ $S$-duality to hold (in the sense that the dual theory is of the same kind as the original theory, albeit with complexified gauge coupling $\tau_D = -1/\tau$) at the quantum level, it suffices that ${\widehat I}' \equiv {\widehat I}'_D$ modulo $2\pi i \mathbb Z$. In other words, the condition for exact $S$-duality to hold in the quantum theory is that the term $2 \pi i \eta \alpha \int_M \delta_D \wedge \delta_D$ must be equal to $2 \pi i \mathbb Z$.

From (\ref{intersection number}), we learn that $\alpha \int_M \delta_D \wedge \delta_D = \alpha D \cap D$ must always be an integer. Therefore, for $2 \pi i \eta \alpha \int_M \delta_D \wedge \delta_D$ to be equal to $2 \pi i \mathbb Z$, the parameter $\eta$ must also be an integer. That is, exact $S$-duality can only be maintained if $\eta$ takes integer values.

\vspace{0.3cm}

\smallskip\noindent{\it Action Under a Shift in Theta-Angle}

Recall also that for a nontrivially-embedded surface operator, one must add the terms $(\alpha^2 / 2) D \cap D$ and $ 2 \pi i \alpha \eta D \cap D$ to (\ref{N}) and (\ref{I_shift}), respectively. Hence, we have in this case
\be
{\widehat I}'_{\Theta} + {\widehat I}'_\eta = i \Theta \left( {\bf N} + (\alpha^2 / 2) D \cap D\right) - 2 \pi i \eta \left(\frak m - \alpha D \cap D \right),
\ee
where $\bf {N}$ is given in (\ref{N}).

For $M$ spin, the variation in ${\widehat I}'_{\Theta}$ under $T: \Theta \to \Theta + 2 \pi$ is now (mod $2 \pi i \mathbb Z$)
\be
\Delta {\widehat I}'_{\Theta} = -2 \pi i \alpha \frak m + \alpha \pi i \mathbb Z,
\ee
where we have made use of the fact that $\alpha D \cap D \in \mathbb Z$.

Suppose we have the transformation
\be
(\alpha, \eta) \to (\alpha, \eta - \alpha)
\label{shift 2}
\ee
under $T:\tau \to \tau + 1$. Then, the corresponding variation in ${\widehat I}'_{\eta}$ will be given by
\be
\Delta {\widehat I}'_{\eta} = 2 \pi i \alpha \frak m - 2  \alpha \pi i \mathbb Z.
\ee

In order for the theory to be invariant under $T: \tau \to \tau +1$ when the parameters of the surface operator transform as in (\ref{shift 2}), we must have $\Delta {\widehat I}'_{\Theta} + \Delta {\widehat I}'_{\eta} = -\alpha \pi i \mathbb Z = 0$ modulo $\ 2\pi i \mathbb Z$. In other words, $\alpha$ can only be even-integer-valued, for $M$ spin.

For $M$ non-spin, the variation in ${\widehat I}'_{\Theta}$ under $T^2: \Theta \to \Theta + 4 \pi$ is now (mod $2 \pi i \mathbb Z$)
\be
\Delta {\widehat I}'_{\Theta} = -4 \pi i \alpha \frak m  +  2\alpha \pi i \mathbb Z.
\ee
Suppose we have the transformation
\be
(\alpha, \eta) \to (\alpha, \eta - 2\alpha)
\label{shift 2 non-spin}
\ee
under $T^2:\tau \to \tau + 2$. Then, the corresponding variation in ${\widehat I}'_{\eta}$ will be given by
\be
\Delta {\widehat I}'_{\eta} = 4 \pi i \alpha \frak m - 4  \alpha \pi i \mathbb Z.
\ee

In order for the theory to be invariant under $T^2: \tau \to \tau +2$ when the parameters of the surface operator transform as in (\ref{shift 2 non-spin}), we must have $\Delta {\widehat I}'_{\Theta} + \Delta {\widehat I}'_{\eta} = - 2 \alpha \pi i \mathbb Z = 0$ modulo $2\pi i \mathbb Z$. In other words, $\alpha$ can only be integer-valued, for $M$ non-spin.

\vspace{0.3cm}

\smallskip\noindent{\it Action Under Overall Duality}

Let us now summarise the action of the transformations $S$, $T$ and $ST^2S$ on the parameters $(\alpha, \eta)$. For
\be
S = \left(\begin{array}{ccc} 0 & & 1 \\ -1 & & 0 \end{array} \right), \qquad T = \left(\begin{array}{ccc} 1 & & 1 \\ 0 & & 1 \end{array} \right) \qquad ST^2S = \left(\begin{array}{ccc} -1 & & 0 \\ 2 & & -1 \end{array}\right),
\ee
We find that $\alpha$ and $\eta$ transform as
\be
(\alpha, \eta) \to (\alpha, \eta) {\frak M}^{-1},
\label{transform2}
\ee
where $\frak M$ is $S$ or $T$ for $M$ spin, or is $S$ or $ST^2S$ for $M$ non-spin. However, in contrast to the previous case of a trivially-embedded surface operator, $\eta$ and $\alpha$ have to be restricted to integer and even-integer values, respectively, for $M$ spin, and only integer values for $M$ non-spin, as explained above.

Recall at this point from $\S$2.1, that $\eta$, by definition, must take values in $\mathbb R / \mathbb Z$. Hence, taking $\eta$ to be integer-valued is equivalent to setting $\eta$ to zero. In other words, exact $S$-duality can only be preserved in the general, twisted ${\cal N}=2$ pure abelian gauge theory for a class of $\textrm{\it{nontrivially-embedded}}$ surface operators which effectively have parameters $(\alpha, \eta) = (\alpha, 0)$. Alternatively, notice that since the term $2 \pi i \eta \mathbb Z$ that results in the non-invariance is a $c$-number independent of the quantum fields, one could instead allow $\eta$ to be non-vanishing and arbitrarily-valued, and claim that exact $S$-duality holds up to a $c$-number only.

Recall from footnote~2 that $\alpha$ (the $S$-dual of $\eta$) takes values in $\mathbb R / \mathbb Z$ too. This means that $\alpha$ also effectively vanishes if it is any integer.  Hence, it will mean that the condition $\Delta {\widehat I}'_{\Theta} + \Delta {\widehat I}'_{\eta} = -\alpha \pi i \mathbb Z = 0 \ \textrm{mod} \ 2\pi i \mathbb Z$, or the condition $\Delta {\widehat I}'_{\Theta} + \Delta {\widehat I}'_{\eta} = -2\alpha \pi i \mathbb Z = 0 \ \textrm{mod} \ 2\pi i \mathbb Z$---which ensures invariance of the theory under $T:\tau \to \tau +1$ or $T^2:\tau \to \tau +2$, respectively---cannot really be satisfied for any non-zero value of $\alpha$. At any rate, the term $\alpha \pi i \mathbb Z $ or $2\alpha \pi i \mathbb Z $, which results in the non-invariance, is a $c$-number independent of the quantum fields. This implies that the symmetry $T: \tau \to \tau +1$ or $T^2: \tau \to \tau +2$ always holds up to a $c$-number only.

In summary, we find that for a nontrivially-embedded surface operator, the parameters $(\alpha, \eta)$ will transform naturally under $SL(2,\mathbb Z)$ (or $\Gamma_0(2)$ when $M$ is non-spin) as shown in (\ref{transform2}). However,   exact $S$-duality holds if and only if $\eta$ is effectively zero, while the overall $SL(2,\mathbb Z)$ (or $\Gamma_0(2)$) duality holds up to a $c$-number at most, regardless.

\vspace{0.2cm}

\newsubsection{Relating the ``Ramified'' and Ordinary Donaldson Invariants}

 We shall now give a brief sketch of how one can, among other things, relate a ``ramified'' analog of the celebrated Donaldson invariants~\cite{Donaldson} to the ordinary Donaldson invariants, using the results we have found above.

Firstly, let us recall some facts about Donaldson-Witten theory (first define in~\cite{TQFT}), which is obtained via twisting an ${\cal N}=2$ pure $SU(2)$ theory in four-dimensions. On a generic, non-zero point in the $u$-plane, where $u = a^2/2$ is the (classical) complex modulus of the theory (which has $R$-charge four under a $U(1)_R$ symmetry, because $a$ has $R$-charge two), the $W^{\pm}$ bosons will be massive. As such, Donaldson-Witten theory would be represented by a topological $U(1)$ theory at low energies or large scales. However, over the special point $u = 1$---where the ``magnetic'' frame is the preferred frame for the $U(1)$ theory---massless monopoles make an appearance~\cite{Seiberg-Witten}. Consequently, the $U(1)$ theory (in the ``magnetic'' frame) will, at $u=1$,  be  coupled to a (twisted) matter hypermultiplet which contains the massless monopoles in question. A similar thing also happens at the point $u=-1$, where massless dyons make an appearance in the effective theory~\cite{Seiberg-Witten}. As a result, the topological observables in Donaldson-Witten theory (which correspond to the ordinary Donaldson invariants) can be related---due to the topological and hence scale-invariance of the theory---to the topological observables of the ``magnetic'' $U(1)$ theory coupled to the massless monopoles and dyons (which correspond to the ordinary Seiberg-Witten invariants~\cite{monopoles}). In particular, one can compute the generating function of the Donaldson invariants in terms of an integral over a generic region of the $u$-plane and the Seiberg-Witten invariants at $u = \pm 1$~\cite{Moore-Witten}. Note that both the topological theories associated to the $u$-plane integral and the Seiberg-Witten invariants are abelian in nature. This just reflects the fact that in the full quantum theory, the $SU(2)$ gauge symmetry is never restored anywhere on the $u$-plane.

Next, note that the ``classical'' surface operator parameter $\alpha$ and its ``quantum'' parameter $\eta$---just like the complexified gauge coupling parameter $\tau(a)$---are expected to receive perturbative and/or non-perturbative quantum corrections. As such, the effective parameters at low-energies will be functions of the original parameters at high-energy. Note also that the assignment of $R$-charge in our theory depends only on the form of its (classical) Lagrangian---which, with the inclusion of surface operators, remains unchanged except that one has to replace $F$ with $F'$ everywhere, and add a charge-free topological term $- i \eta (\delta_D \wedge F')$. As such, the $R$-charges of all fields in a ``ramified'' extension of the theory will be the same as before. In particular, the (``ramified'') gauge field and scalar $a$ continue to have $R$-charges zero and two, respectively. In addition, since the $SU(2)$ gauge symmetry is never restored anywhere on the $u$-plane in the full quantum theory, the specific reduction of the $SU(2)$ gauge group to its $U(1)$ maximal torus along $D$ whenever a surface operator is present, is irrelevant in regards to computations on the $u$-plane. Last but not least, because the effective low-energy Lagrangian $L'$ and its shift $ \delta L' = -{\gamma \over \pi^2} F' \wedge F'$ (with constant $\gamma$) due to the chiral anomaly take the same form (albeit with $F$ replaced by $F'$ everywhere) as in the case without surface operators, and because the effective instanton number $k$ that appears in the instanton factor $e^{-8\pi^2 k / g}$ is again an integer,\footnote{The assertion that the effective instanton number $k$ continues to be an integer in the presence of a surface operator, is a subtle point which can be justified as follows. Firstly, the instanton number is, in this case, given by $- {1 \over 8 \pi^2} \int_M \textrm{Tr} F' \wedge F' = - {1 \over 8 \pi^2} \int_M \textrm{Tr} F \wedge F + \textrm{Tr} \alpha \frak m - {1\over 2} \textrm{Tr} \alpha^2 D \cap D$. The first term $-{1 \over 8 \pi^2} \int_M \textrm{Tr} F \wedge F$ is always an integer while the last term $-{1\over 2} \textrm{Tr} \alpha^2 D \cap D$ is also an integer by virtue of the condition (\ref{intersection number 2}). The second term $\textrm{Tr} \alpha \frak m$ however, is not an integer, since $\alpha \in {\frak t} / \Lambda_{\textrm{cochar}}$~\cite{Gukov-Witten}. Nevertheless, since ${\widehat I}'_{\eta} = -2 \pi i \textrm{Tr}\eta\frak m + \dots$, one can absorb this second term by a shift in $\eta$ when computing the overall instanton factor. Thus, the $\it{effective}$ instanton number will be an integer as stated.} the exact form of the holomorphic prepotential ${\cal F}(A)$---which determines the exact expression for $\tau(a)$ and the asymptotic behaviour of $a$ and $a_D$ etc.---should remain unchanged in the presence of surface operators. Altogether, this implies that apart from the fact that the parameters of a surface operator might be scale-dependent, the main features of the ordinary theory without surface operators should carry over to the generalised theory with surface operators.

Now, consider Donaldson-Witten theory $\it{with}$ surface operators, such that its topological observables would correspond to a ``ramified'' analog of the ordinary Donaldson invariants. Based on what we have said above, we deduce that over a generic region of the $u$-plane, the effective topological abelian action would be given by ${\widehat I}'$. At $u= 1$, the effective topological abelian action would be given by ${\widehat I}'_D$ coupled to a (twisted) matter hypermultiplet which contains the massless monopoles of interest. Likewise at $u=-1$, the effective topological action would involve a coupling to massless dyons instead. The topological observables of the effective theories at $u= \pm 1$, should then correspond to a ``ramified'' analog of the ordinary Seiberg-Witten invariants. For trivially-embedded surface operators, one might---since (exact) $S$-duality always holds in the  abelian theory with action ${\widehat I}'$---be able to compute a generating function of ``ramified'' Donaldson invariants in terms of an integral over a generic region of the $u$-plane, and the ``ramified'' Seiberg-Witten invariants at $u = \pm 1$.

On the other hand, recall that for nontrivially-embedded surface operators, (exact) $S$-duality of the abelian theory  with action ${\widehat I}'$ holds if and only if the parameters of the surface operators take the form $(\alpha,\eta) = (\alpha, 0)$. Since the surface operators of the abelian theory in the ``magnetic'' frame (with partial action ${\widehat I}'_D$) at $u = 1$ will have parameters $(0, -\alpha)$ as such, it would mean that there is no ``ramification'' at $u=1$, that is, one will end up computing the $\it{ordinary}$ Seiberg-Witten invariants at $u=1$. This means that if we start with a high-energy ``quantum'' parameter that vanishes, and make a judicious choice of the remaining high-energy ``classical'' parameter for which the corresponding low-energy ``quantum'' parameter also vanishes, one can express the generating function of ``ramified'' Donaldson invariants in terms of an integral over a generic region of the $u$-plane and the ordinary Seiberg-Witten invariants at $u = \pm 1$ (since the contributions at $u=-1$ will be related by a discrete symmetry to those at $u=1$).

For $b^+_2 > 1$, where $b^+_2$ is the self-dual second Betti number of $M$, the contribution from the $u$-plane integral vanishes, because there are too many fermionic zero-modes in the integral measure that cannot be completely absorbed by bringing down interaction terms in the integrand. Hence, in the nontrivially-embedded case, and for the appropriate choices of the high-energy parameters as mentioned above, the ``$\it{ramified}$'' Donaldson invariants will be expressed solely in terms of the ordinary Seiberg-Witten invariants, which, in turn, must be expressed solely in terms of the $\it{ordinary}$ Donaldson invariants, since $b^+_2 > 1$. This physical observation happens to be consistent with a remarkable mathematical result of Kronheimer and Mrowka~\cite{KM}, which for $b^+_2 > 1$ and $D\cap D \neq 0$, expresses the ``ramified'' analog of the Donaldson invariants purely in terms of the ordinary Donaldson invariants. This result is central to the proof of a structure theorem by Kronheimer and Mrowka~\cite{KM,RR}, that provides one with a basis for interpreting the ordinary Donaldson invariants in terms of the ordinary Seiberg-Witten invariants, as, was done, in~\cite{monopoles}.

A detailed analysis of the above matters, and more, will appear elsewhere in forthcoming work~\cite{MC}.

\newsubsection{Relation to the Pure ${\cal N}=4$ Theory}

Before we end this section, let us comment on the relation between our pure ${\cal N} =2$ theory and the pure ${\cal N} =4$ theory (whose twisted version has been considered in~\cite{Gukov-Witten}).  

It is known that the pure ${\cal N} = 4$ theory in question can be obtained by including hypermutliplet matter in the adjoint representation to the pure ${\cal N}=2$ theory considered above with prepotential ${\cal F}(A) = {1 \over 2} \tau_{cl} A^2$.  Since the resulting Lagrangian is now ${\cal N} =4$ supersymmetric, the supersymmetric field configurations that one obtains by setting the supersymmetric variations of the fermions to zero, will be different; they will involve, in addition to the field strength $F$, the Higgs field $\phi$. (For example, in a particular topological twist of the resulting ${\cal N} =4$ theory on $M = D \times D'$, where $D'$ is much smaller than $D$, these configurations are given by eqn.~(2.1) of~\cite{Gukov-Witten}.) Consequently, as mentioned earlier, a supersymmetric surface operator in the ${\cal N}=4$ theory is one that will introduce a singularity in $\it{both}$ $F$ and $\phi$ along $D$ (see eqn.~(2.2) of~\cite{Gukov-Witten} for an example); this is in contrast to the ${\cal N}=2$ case, which only involves a singularity in $F$.  As such, one would also need to shift $\phi$ appropriately in the (Euclidean) Lagrangian in order to have a non-vanishing contribution to the path-integral, since the positive-definite kinetic term of $\phi$ would otherwise be singular upon integration over $M$.  

At any rate, one can, at least in the abelian case, proceed to prove explicitly that the parameters of the surface operator do transformation naturally under the $SL(2,\mathbb Z)$ duality of the ${\cal N} =4$ theory. This has been done for the twisted theory relevant to the ``ramified'' geometric Langlands program in $\S$2.4 of~\cite{Gukov-Witten}, and the somewhat ad-hoc approach taken there is tantamount to our analysis in $\S$2.2-2.3 if we add to our Lagrangian (\ref{I'}) the appropriate massless ${\cal N}=2$ adjoint hypermultiplet terms that are defined  to map---in the sense of (\ref{equal})---to themselves in the $\it{dual}$ fields under $S$-duality. The transformation of $(\alpha, \eta)$---which involves the term (\ref{I'_2})---is proved as in $\S$2.2-2.3, while the transformation of the (classical) parameters $(\beta, \gamma)$ which characterise the singularity of the  Higgs field $\phi$ along $D$ (see eqn.~(2.2) of~\cite{Gukov-Witten}), is obtained by a comparison of the kinetic energy of $\phi$ in the original and dual theory;  the parameters $(\beta, \gamma)$ ought to transform in accordance with the fact that the kinetic energy is invariant under $S$-duality.   

For brevity, we shall not elaborate on this further, except to comment that our above-described framework can also be applied to an arbitrary prepotential ${\cal F}(A)$; we need not restrict ourselves to ${\cal F}(A) = {1 \over 2} \tau_{cl} A^2$ as mentioned above, which is required only to relate the ${\cal N}=2$ theory to the standard pure ${\cal N}=4$ theory.

\newsection{Effective Interaction on the $u$-Plane}

In all our discussions so far, we have implicitly assumed that $S$-duality would hold in the full quantum theory as long as it holds at the level of the quantum action. This however, is a naive assumption, because even though the allowable operators in the correlation functions must be duality-invariant, the measure of the path-integral will transform nontrivially under the transformations that map the abelian theory in the ``electric'' frame to its supposedly equivalent theory in the ``magnetic'' frame.

Nevertheless, it was shown in~\cite{SO abelian} that for the non-supersymmetric counterpart of our theory, the modular anomaly of the partition function manifests itself only when the four-manifold is curved. This implies that whenever $M$ is a curved four-manifold, the part of the measure involving the gauge field will transform nontrivially under $S$-duality. Thus, we are bound to have a modular anomaly in the theory on curved $M$ (as presented so far).

On the other hand, if $M$ is flat, there will not be any modular anomaly due to the transformation of the gauge field. However, one could potentially have an anomaly due to the nontrivial transformation of the fermions in the measure. In such a case, a little thought would reveal that the computation of the modular anomaly should be similar to a computation of an $R$-symmetry anomaly (which is also due to a nontrivial transformation of fermions of the same nature in the measure). This implies that the modular anomaly should be quantified by the index of the kinetic operator acting on the fermions in the Langragian, which, typically involves the Euler characteristic $\chi$ and signature $\sigma$ of $M$, that, in turn will also vanish if $M$ is flat. This assertion will be justified shortly.

Therefore, the issue is really when $M$ is curved. However, when $M$ is curved in the twisted abelian theory at hand, interaction terms proportional to $\int_M f(u) \ \textrm{tr} R \wedge \widetilde R$ or $\int_M g(u) \ \textrm{tr} R \wedge R$ become possible, where $\textrm{tr} R \wedge \widetilde R$ and $\textrm{tr} R \wedge R$ are the densities whose integrals are proportional to $\chi$ and $\sigma$, respectively, and $f$ and $g$ are holomorphic functions of $u$.\footnote{These interaction terms result from integrating out the massive $SU(2)$ partners of the light fields in the low-energy theory whose Lagrangian is given by (a twisted version of) $L'$. The reason why these terms are admissible as interaction terms of the twisted (and hence topological and BRST-invariant) abelian theory when $M$ is curved, is because $f$ and $g$ are BRST-invariant while $\textrm{tr} R \wedge \widetilde R$  and $\textrm{tr} R \wedge R$ are locally constructed functions of the metric related to the topological invariants $\chi$ and $\sigma$ of $M$. These terms would vanish when $M$ is flat.} This implies that one should be able to find---through these additional interaction terms---an $S$-dual extension of the twisted abelian theory when $M$ is curved.

In this section, we shall determine the factor
\be
\textrm{exp} \left(b(u) \chi + c(u) \sigma \right)
\label{factor}
\ee
which appears (for $u$ constant) in the path-integral due to these interaction terms. The reason for doing so is that we would like to affirm our above claim that with these interaction terms, a consistent $S$-dual extension of the twisted abelian theory exists for $\it{any}$ smooth $M$. Moreover, one would also need to know this factor in order to compute the exact expression of the $u$-plane integral. To this end, we shall generalise the analysis in $\S$3 of~\cite{abelian S-duality} to include surface operators embedded in $M$.

\newsubsection{Asymptotic Behaviour}

A physically consistent choice of the functions $b(u)$ and $c(u)$ would ensure that the factor in (\ref{factor}) matches the expected asymptotic behaviour of the theory at the special points $u=\infty$, $1$ or $-1$, where one is in a region of weak coupling and  where massless monopoles or dyons make their appearance, respectively.

Let us first consider the behaviour of the abelian theory at large $u$, where the ``electric'' frame with scalar field $a$ is the preferred frame of the theory. Note that the microscopic, topological ${\cal N} = 2$ pure $SU(2)$ theory associated to our abelian theory in the ``electric'' frame has an anomalous $U(1)_R$ symmetry, and the anomaly is quantified by the index of the kinetic operator acting on the fermions in the non-abelian Lagrangian. Recall at this point that as explained earlier, the connections in the Lagrangian which contribute non-vanishingly to the path-integral are such that their corresponding field strengths take the form $F' = F - 2\pi\alpha \delta_D$. In other words, the various dynamical fields in the theory are effectively coupled to the gauge field $A'$, whose field strength is $F' = dA' + A' \wedge A'$.\footnote{The prime in $A'$ and $F'$ means that they correspond to a connection and curvature of an $SU(2)$-bundle over $M$ that is equivalent to the $SU(2)$-bundle over $M \setminus  D$ with singular connection $A$ and curvature $F$ along $D$.} Consequently, the index would be given by the (virtual) dimension $\textrm{dim}({{\cal M}}_{\textrm{inst}})$ of the moduli space ${{\cal M}}_{\textrm{inst}}$  of $SU(2)$-instantons associated with $A'$; that is, the anomaly would be given by
\be
\Delta R = 8k' - {3 \over 2} (\chi + \sigma),
\label{deltaR}
\ee
where $k' = - {1\over 8 \pi^2} \int_M \textrm{Tr} F' \wedge F'$ is the corresponding instanton number. (That is, operators with non-zero expectation value must have $R$-charge equal to $\Delta R$.)

The term of interest here is the part that involves the coupling to gravity, namely,
\be
- {3\over 2} (\chi + \sigma).
\label{anomaly SU(2)}
\ee
(For flat $M$, this term will vanish, as highlighted earlier.)

For large $u$, the $W^{\pm}$ bosons will be extremely massive, and the $SU(2)$ gauge symmetry will be spontaneously broken down to a $U(1)$ gauge symmetry, whereby we would have a $U(1)$ theory with (a twisted version of the) Lagrangian $L'$ of (\ref{L'}). The zero-modes of the fermion fields $\lambda$ and $\bar\lambda$ in (the twisted version of) $L'$ that live in the $u$ multiplet, would carry an anomaly equal to the index of the kinetic operator acting on them in the Lagrangian. Since the fermion fields are not coupled to the gauge field in the kinetic terms of the abelian theory, the index would be the same as in the case without surface operators; that is, the anomaly carried by the fermion fields will be given by $-(\chi + \sigma)/2$. The remaining anomaly $-(\chi + \sigma)$ must manifest in an interaction that results from integrating out the massive $SU(2)$ partners of the light fields. Since the interaction must not have any derivatives (else it would vanish if $u$ is constant), it will be of the forrm (\ref{factor}). Thus, at large $u$, since $u$ has $R$-charge four, we must have
\be
e^{b\chi + c\sigma} \sim u^{(\chi + \sigma)/4} \quad \textrm{for} \quad u \to \infty.
\label{ularge}
\ee
Notice that the factor has $R$-charge $(\chi + \sigma)$ and not $-(\chi + \sigma)$. This is because the factor results from integrating out the massive modes in the measure, and is hence $\it{not}$ an operator of the theory. Therefore, its $R$-charge must be such that when one inserts an actual operator with $R$-charge equal to the remaining anomaly of $-(\chi + \sigma)$ in the path-integral, the total $R$-charge is zero, so that the resulting correlation function will be invariant under the $U(1)_R$ symmetry, as required.

Let us now analyse the behaviour near $u = \pm 1$. The effective Seiberg-Witten theory at $u=1$ has an (accidental) low-energy $U(1)_R$ symmetry. The anomaly of this symmetry is again given by the index of the appropriate kinetic operator in the effective Lagrangian at $u=1$. For a trivially-embedded surface operator, the index would be given by the dimension $\textrm{dim}({\cal M}_{\textrm{SW}})$ of the moduli space ${\cal M}_{\textrm{SW}}$ of solutions to the ``ramified'' Seiberg-Witten equations.\footnote{The ``ramified'' Seiberg-Witten equations are given by $(F_D - 2\pi \eta \delta_D)^+ + i(\overline {\cal M} {\cal M})^+ = 0$ and $D {\cal M} = 0$, where the monopole $\cal M$ is a complex spinor field obtained via twisting the scalar fields in the massless matter hypermultiplet that is coupled to ${\widehat I}'_D$ at $u=1$, $D$ is the Dirac operator, and the ``$+$'' superscript indicates the self-dual part of the respective fields.}  Thus, the anomaly would, in this case, be given by
\be
-{1\over 2}(\chi + \sigma) + {c_1({\tilde L}')^2} - {\sigma \over 4},
\label{3.5}
\ee
where ${\tilde L}'$ is the $U(1)$-bundle associated to the dual, ``magnetic'' photon with curvature field strength $F'_D = F_D - 2 \pi \eta \delta_D$.\footnote{To see this, first note that  the dimension of the moduli space of the ordinary Seiberg-Witten equations will be given by the index of an elliptic operator $T$~\cite{monopoles}. $T$ is defined by $T: \Lambda^1 \oplus (S^+ \otimes {\tilde L}) \to \Lambda^0 \oplus\Lambda^{2,+} \oplus (S^- \otimes {\tilde L})$, where $\Lambda^p$ denotes the bundle of real-valued $p$-forms on $M$, $S^{\pm}$ are positive and negative chirality spinor bundles on $M$, and ${\tilde L}$ is a complex line bundle whose curvature is an ordinary, non-singular field strength of the dual, ``magnetic'' photon. If we now include surface operators, the Seiberg-Witten equations would be modified as stated in footnote~8; in particular, we would now have $(F_D - 2\pi \eta \delta_D)^+ = -i(\overline {\cal M} {\cal M})^+$. This just corresponds to the fact that $\cal M$ (or $\overline {\cal M}$) is now a section of the bundle $S^+ \otimes {\tilde L}'$ (or $S^+ \otimes {\tilde L}^{'-1}$), where ${\tilde L}'$ is the  $U(1)$-bundle associated to the dual, ``magnetic'' photon with curvature field strength $F'_D = F_D - 2 \pi \eta \delta_D$. Hence, the expression for the index with surface operators  is simply the expression for the ordinary index but with $\tilde L$ replaced by $\tilde L'$, as indicated in (\ref{3.5}).}  For the case of a nontrivially-embedded surface operator, since there is $\it{no}$ ``ramification'' in  the preferred ``magnetic'' frame near $u=1$ as explained earlier, one just needs to replace ${\tilde L}'$ in (\ref{3.5}) with the ordinary $U(1)$-bundle $\tilde L$ that has non-singular curvature field strength. As usual, $-(\chi + \sigma)/2$ will be the anomaly carried by the fermions $\lambda_D$ and $\bar \lambda_D$ in (the twisted version of) the Lagrangian $L'_D$ that are related to $u$ by duality and supersymmetry. The ${c_1(\tilde L')^2} - {\sigma / 4}$ part is due to the monopoles that become massless at $u=1$. Since the contribution ${c_1(\tilde L')^2}$ appears in the behaviour of the effective gauge couplings near $u=1$~\cite{Seiberg-Witten}, it would mean that the gravitational anomaly $- {\sigma / 4}$ ought to manifest itself in an interaction of the form (\ref{factor}) when one integrates out the light monopoles near $u=1$. As mentioned, near $u=1$, the ``magnetic'' frame in the variable $a_D$ is the preferred frame. In addition, the mass of the light monopoles near $u=1$ is proportional to $a_D$. Altogether, this implies that $a_D \approx c_0(u-1)$, where $c_0$ is a constant~\cite{Seiberg-Witten}. In other words, $(u-1)$ has $R$-charge two (since $a_D$, like $a$, has $R$-charge two) near $u=1$. Hence, this means that
\be
e^{b\chi + c\sigma} \sim (u-1)^{\sigma/8} \quad \textrm{for} \quad u \to 1.
\label{u1}
\ee
By a similar argument, from the behaviour of the theory near $u=-1$, where an interaction of the form (\ref{factor})---that carries the gravitational $U(1)_R$ symmetry anomaly $- {\sigma / 4}$---results from integrating out light dyons that become massless only at $u=-1$, we find that
\be
e^{b\chi + c\sigma} \sim (u+1)^{\sigma/8} \quad \textrm{for} \quad u \to -1.
\label{u-1}
\ee

From the asymptotic conditions (\ref{ularge}), (\ref{u1}) and (\ref{u-1}), we conclude that
\be
e^{c\sigma} \sim (u^2-1)^{\sigma/8}.
\ee
However, there is no holomorphic ``function'' $e^{b\chi}$ in $u$ that can satisfy the conditions (\ref{ularge}), (\ref{u1}) and (\ref{u-1}) all at once. Nonetheless, we should be able to determine the correct form of $e^{b\chi}$ once  we know the conditions for which $S$-duality is preserved even when $M$ is curved. This will be done in the next two subsections.

\newsubsection{Modular Anomaly}

We shall now discuss the modular anomaly of the theory when the four-manifold $M$ is curved. The analysis will be the same for any embedding of surface operators. As such, we shall not need to discuss the trivially and nontrivially embedded surface operator separately.

To proceed, first recall that the physically equivalent actions ${\widehat I}'$ and ${\widehat I}'_D$ are such that they map exactly into each other under the $S$-duality transformations. Next, note that operators defined in the correlation functions are, at the outset, defined to be completely gauge and duality-invariant. Consequently, the only source of a modular anomaly comes from the way the path-integral measure transforms under $S$-duality.

In order to ascertain the modular anomaly, we will need to analyse how the integration measure of the scalar fields,  fermions, and gauge fields transform under $S$-duality. We will analyse them in turn now.

\smallskip\noindent{\it Integration Measure of the Scalars}

Notice from the explicit expression of $L'$ in (\ref{L'}) that the kinetic energy of $a$ and $\bar a$ (in the twisted theory) is proportional to $\textrm{Im} \tau$. Thus, the integration measure for every mode $a'$ and ${\bar a}'$ of the scalar fields $a$ and $\bar a$ will take the form
\be
(\textrm{Im} \tau) da' d {\bar a}'.
\ee

From the equivalence of $L'$ and $L'_D$ in (\ref{L'}) and (\ref{Dual L}), we have $a_D = \tau a$ and ${\bar a}_D = \bar \tau \bar a$. Since $\tau_D = - {1 / \tau}$ and $(\textrm{Im}\tau_D) = (\textrm{Im} \tau) / \tau \bar\tau$, we find that
\be
(\textrm{Im} \tau) da' d\bar a' = (\textrm{Im} \tau_D) da'_D d{\bar a}'_D.
\ee
Thus, the integration measure of the scalar fields is completely duality-invariant.

\smallskip\noindent{\it Integration Measure of the Fermions}

From the explicit expression of $L'$ in (\ref{L'}), we see that the kinetic energy of the fermions $\lambda$ and $\bar\lambda$ (in the twisted theory) is also proportional to $\textrm{Im} \tau$. This means that the modes of  $\lambda$ and $\bar\lambda$ take the forms $\sqrt {\textrm{Im}\tau}\ \xi$ and $\sqrt {\textrm{Im}\tau}\ \bar\xi$, respectively, where $\xi$ and $\bar \xi$ are $\it{normalised}$ fermi modes. Since the modes of the fermions are $\it{Grassmannian}$ in nature, the fermionic integration measure for each of these modes can be written as
\be
d(\sqrt {\textrm{Im}\tau}\ \xi) = {d \xi \over \sqrt {\textrm{Im}\tau}} \qquad \textrm{and} \qquad d(\sqrt {\textrm{Im}\tau}\ \bar\xi) = {d \bar\xi \over \sqrt {\textrm{Im}\tau}}.
\ee

From the equivalence of $L'$ and $L'_D$ in (\ref{L'}) and (\ref{Dual L}), we have $\lambda_D = \tau \lambda$ and ${\bar \lambda}_D = \bar \tau \bar \lambda$. Given that $\tau_D = - 1 /\tau$, and $(\textrm{Im}\tau_D) = (\textrm{Im} \tau) / \tau \bar\tau$, we find that
\be
{d\xi \over \sqrt{\textrm{Im}\tau}} = {\sqrt{\tau \over \bar \tau}} {d \xi_D \over \sqrt{\textrm{Im}\tau_D}} \qquad \textrm{and} \qquad {d\bar\xi \over \sqrt{\textrm{Im}\tau}} = {\sqrt{\bar\tau \over \tau}} {d \bar\xi_D \over \sqrt{\textrm{Im}\tau_D}}.
\ee

Let us denote the fermionic measure in the ``electric'' and ``magnetic'' frames as $d\mu^F$ and $d\mu^F_D$. In order to ascertain how $d\mu^F$ and $d\mu^F_D$ are related under a duality transformation, first note that there is a pairing of non-zero modes of $\lambda$ and $\bar\lambda$. As a result, the non-zero cancelation of the factors $\sqrt{\tau / \bar \tau}$ and $\sqrt{\bar \tau /\tau}$ in the total measure $d\mu^F$, depends only on the difference between the number of zero-modes of $\lambda$ and $\bar\lambda$. Next, recall that this difference is given---via the relevant index theorem---by $-{(\chi + \sigma)/ 2}$. Altogether, this means that
\be
d\mu^F = \tau^{-{(\chi + \sigma)/ 4}} {\bar\tau}^{{(\chi + \sigma)/ 4}}  d\mu^F_D.
\ee

\smallskip\noindent{\it Integration Measure of the Gauge Fields}

Last but not least, we come to the measure of the gauge fields. In order to determine how the measure will transform under the duality transformations, first note that from the gauge kinetic and theta terms in $L'$ of (\ref{L'}), we know that the gauge kinetic and theta terms of the topological action ${\widehat I}'$ appear as
\be
{\widehat I}' = {i\bar\tau\over 4 \pi} \int_M {F'}^+ \wedge \star {F'}^+ - {i\tau\over 4 \pi} \int_M {F'}^- \wedge \star {F'}^- - i\eta \int_D F' + \cdots,
\label{gauge field part}
\ee
where the ``$+$'' and ``$-$'' superscripts indicate the self-dual and anti-self-dual parts of $F'$, respectively. Since the transformation of the gauge field measure should be independent of the explicit form of $\tau$, let us, for simplicity, pick $\tau$ such that $d\tau/da = d^2\tau /da^2 = 0$. Then, from $L'$ in (\ref{L'}), it is clear that the dependence of ${\widehat I}'$ on the field strength $F'$ is $\it{completely}$ captured by (\ref{gauge field part}), that is, there is now no $F'$ dependence in the terms represented by the ellipsis. As such, we can read across from~\cite{SO abelian} (which studied the non-supersymmetric counterpart to ${\widehat I}'$ without the terms represented by the ellipsis) to determine the transformation of the gauge field measure under a duality transformation. The result is\footnote{Due to a sign difference in the theta-term of ${\widehat I}'$ relative to that in~\cite{SO abelian}, one must, in what follows, make the swop $\tau\leftrightarrow -\bar\tau$ in reading across the result from~\cite{SO abelian}.}
\be
d\mu^G =  \tau^{-{(\chi-\sigma)/ 4}} {\bar\tau}^{-{(\chi+\sigma)/ 4}} d\mu^G_D,
\ee
where strictly speaking, $d\mu^G$ and $d\mu^G_D$ should be interpreted as the gauge field partition function of the ``electric'' and ``magnetic'' theories, respectively.

\smallskip\noindent{\it The Resulting Modular Anomaly}

Let $d\mu$ and $d\mu_D$ denote the total integration measure in the ``electric'' and ``magnetic'' frames of the theory. Then, by putting the separate integration measures for the scalars, fermions and gauge fields found above together, we get
\be
d\mu = \tau^{-\chi / 2}  d\mu_D.
\label{total measure tx}
\ee
Hence, we see a hint that the factor $e^{b\chi}$ ought to be such that it cancels the factor $\tau^{-\chi / 2}$ so as to maintain $S$-duality on any $M$.

\newsubsection{Final Determination}

We shall now determine the explicit form of $e^{b\chi}$ and hence, the effective interaction $e^{b\chi + c \sigma}$ on the $u$-plane.

If the theory is to be completely $S$-dual, from our discussion so far, it would mean that we must have
\be
 e^{b\chi} \ d\mu = e^{b_D\chi} \ d\mu_D.
\ee
From (\ref{total measure tx}), this means that
\be
e^{b_D\chi} =  \tau^{-\chi / 2} \ e^{b\chi},
\label{eb tx}
\ee
that is, $e^{b\chi}$ is a holomorphic ``function'' in $u$ that must transform with modular weight $-\chi/2$ under $S$-duality.

A little thought reveals that such a ``function'' can be written as
\be
e^{b\chi} = \left( (u^2 -1) {d\tau \over du} \right)^{\chi/4}.
\ee
Therefore, we should also have
\be
e^{b_D\chi} = \left( (u^2 -1) {d\tau_D \over du} \right)^{\chi/4}.
\ee
From the relation $\tau_D = -1/\tau$, we find that $e^{b\chi}$ indeed satisfies (\ref{eb tx}). Also, for $u \to \infty$, $a_D \approx  i a \textrm{ln} u / \pi$~\cite{Seiberg-Witten}. Since $\tau = da_D / da$, we have $d\tau/du \approx 1/u$. So $e^{b\chi} \sim u^{\chi /4}$ for $u \to \infty$, as required of condition (\ref{ularge}). Last but not least, we must also check that the ``zero'' in (\ref{u1}) as one approaches $u=1$, is not violated with the above choices of $e^{b\chi}$ and $e^{b_D\chi}$. To proceed, recall that near $u=1$, the ``magnetic'' frame is preferred. Thus, we should check that $e^{b_D\chi}$ converges to a non-vanishing constant near $u=1$. Near $u=1$, we have $d\tau_D / du \sim 1/(u-1)$~\cite{Seiberg-Witten}. Hence, $e^{b_D\chi} \to 2^{\chi/4} \in \mathbb R^+$ when $u \to 1$, as required. Likewise, it can be shown that the ``zero'' in (\ref{u-1}) as $u \to -1$ will be preserved for the choices of $e^{b\chi}$ and $e^{b_D\chi}$ above. Therefore, like the factor $e^{c \sigma}$ determined earliier, $e^{b \chi}$ has the required asymptotic behaviour near $u = \infty$, $1$ and $-1$. Hence, we conclude that it must be of the correct form.

In summary, this means that in any computation on the $u$-plane, one must include in the path-integral the interaction factor
\be
e^{b\chi + c \sigma} = \left( (u^2-1) {d\tau \over du}\right)^{\chi/4} (u^2 -1)^{\sigma/8}.
\label{interaction}
\ee
Despite some important differences in the theories with and without (arbitrarily embedded) surface operators, we find---after a careful analysis of the matter---that the interaction factor (\ref{interaction}) is identical to the one computed in~\cite{abelian S-duality} for the ``unramified'' case. In hindsight, this is not surprising: the additional interaction results from integrating out modes of fields that appear massive in the low-energy theory and is thus, gravitational in nature, while the parameters that characterise a surface operator are  $c$-numbers (valued in some Lie algebra) that do not feature in any computation involving gravity (unless the theory is gauge field-coupled to (super)gravity).

\newsection{Dependence on Second Stiefel-Whitney Classes, a $\textrm{Spin}^c$ Structure and an Interesting Phase Factor}

In this final section, we shall, in anticipation of forthcoming work~\cite{MC} that will compute the generating function of ``ramified'' Donaldson invariants in terms of a $u$-plane integral and the ``ramified'' Seiberg-Witten invariants at $u = \pm 1$, investigate the possible dependence of the low-energy abelian theory (and consequently, the Seiberg-Witten theory)  on certain second Stiefel-Whitney classes. For non-spin manifolds, we shall also investigate the possible appearance of a $\textrm{Spin}^c$ structure. In doing so, we will stumble upon an interesting phase factor---which does not necessarily vanish even on spin manifolds---that is absent in the ``unramified'' case. To this end, we shall generalise the analysis in $\S$4 of~\cite{abelian S-duality} to include surface operators, and consider the underlying microscopic gauge groups to be either $SU(2)$ or $SO(3)$.

\vspace{-0.0cm}
\smallskip\noindent{\it The Mathematics of Spontaneously Broken Gauge Symmetry}

However, before we can proceed any further, we will need to understand the mathematical interpretation of the spontaneous breaking of the underlying $SU(2)$ or $SO(3)$ gauge symmetry to its $U(1)$ subgroup on a generic point in the $u$-plane,  at low energies.

For such gauge groups mentioned herein, their connections are said to be reducible~\cite{Marcos}. For example, take $V$ to be the rank two $SU(2)$-bundle. At low energies, it has a decomposition $V = {T} \oplus {T}^{-1}$, where $T$ is a complex line bundle with connection $C$ that corresponds to the unbroken $U(1)$ on a generic point in the $u$-plane. The low-energy gauge field is thus given by $C$.

Similarly, the rank three $SO(3)$-bundle $E$ can be decomposed at low energies to $E = {\bf R} \oplus {\cal L}$, where $\bf R$ denotes a trivial rank one real bundle over $M$, and $\cal L$ is a complex line bundle that corresponds to the unbroken $U(1)$. The connection on $\cal L$ is therefore the low-energy gauge field $\cal A$. If $w_2(E) = 0$, $E$ can be lifted to the $SU(2)$-bundle $V$, such that ${\cal L} = {T}^{\otimes 2}$ and ${\cal A} = 2{C}$.

\vspace{-0.0cm}
\smallskip\noindent{\it A Topological Constraint}

 When $E$ exists and $w_2(E) = 0$, the first Chern class $c_1(\cal L)$ is an integral cohomology class. This implies the Dirac quantisation condition $\int_U F = 2 \pi \mathbb Z$, where $F = d \cal A$ is the unbroken $U(1)$ field strength associated to the microscopic $SO(3)$ gauge group, and $U$ is a two-cycle in $M$.

However, when $w_2(E) \neq 0$,  $c_1(\cal L)$  is a class that lives in the refined lattice~\cite{Marcos}
\be
c_1({\cal L}) \in  2 H^2(M, \mathbb Z) + w_2(E).
\label{c1}
\ee

For simplicity, let us assume that there is no torsion in the cohomology of $M$, so that we can pick a basis of two-dimensional cycles $U_\gamma$ in $H_2(M,\mathbb Z)$. Then, $w_2(E)$ can be described by the condition
\be
\int_{U_\gamma} w_2(E) = c_\gamma,
\label{stiefel-whitney}
\ee
where $c_\gamma = 0$ or $1$.

From (\ref{c1}), we find the (refined) Dirac quantisation condition for the field strength $F = d \cal A$ to be
\be
\int_{U_\gamma} F  = 2 \pi (2 k_\gamma + c_\gamma),
\label{Dirac quantisation}
\ee
where $k_\gamma \in \mathbb Z$.\footnote{Since $F$ is the ``ramified'' field strength, the integral over $U_\gamma$---if $D$ coincides with $U_\gamma$---will depend on the extension of the $U(1)$-bundle (with curvature $F$) over $D$. Nevertheless, the topological condition of (\ref{Dirac quantisation}) will continue to hold.}

\newsubsection{Dependence on $w_2(E)$}

 Let us take the gauge group to be $SO(3)$ instead of $SU(2)$ in the ``unramified'' Donaldson-Witten theory. Then, it is known that the generating functions of Donaldson invariants depend on the second Stiefel-Whitney class $w_2(E)$ of the rank three $SO(3)$-bundle $E$~\cite{Donaldson}. This has also been physically demonstrated in~\cite{Moore-Witten} through the expression of the generating function for Donaldson invariants in terms of the contributions from the $u$-plane integral and the Seiberg-Witten invariants at $u =\pm 1$. Naturally, one would expect a similar dependence in the ``ramified'' version of the story. Let us investigate this further.

 \vspace{0.0cm}
 \smallskip\noindent{\it Abelian Duality in Low-Energy $SO(3)$ Gauge Theory}

As mentioned above, the explicit dependence on $w_2(E)$ in the ``unramified'' case can also be seen from the contributions at $u = \pm 1$, that is, from the abelian theory in the ``magnetic'' frame. As such, we shall determine if the $\it{dual}$, ``magnetic'' abelian theory will depend on $w_2(E)$ in the ``ramified'' case.

In order to do so, it is clear that because the curvature of $E$ is just the field strength of the gauge field, it suffices to analyse only the gauge field dependent part of the action. To this end, first note that since our analysis would again be independent of the explicit form of $\tau$, let us, for simplicity, choose $\tau$ such that $d \tau / da =0$. Then, the gauge field dependent part of the topological action ${\widehat I}'$ of the abelian ``electric'' theory will be given by
\be
{\widehat I}'_{gauge}({\cal A}, \tau, \alpha, \eta) = {i\bar\tau\over 4 \pi} \int_M {F'}^+ \wedge \star {F'}^+ - {i\tau\over 4 \pi} \int_M {F'}^- \wedge \star {F'}^- - i\eta \int_D F'.
\label{action gauge field}
\ee
Note that ${\widehat I}'_{gauge}$ takes the same form as the action considered in~\cite{SO abelian}. Hence, we shall adopt the approach taken in~\cite{SO abelian} toward our analysis of ${\widehat I}'_{gauge}$ at hand.

\vspace{0.2cm}
\hspace{-1.0cm}(i). \emph{An Extended Gauge Symmetry}

Firstly, let us introduce a two-form $\bf{g}$ that is invariant under the ordinary abelian gauge symmetry ${\cal A} \to {\cal A} -d\epsilon$ (where $\epsilon$ is a zero-form). We would then like to define the following extended gauge symmetry
\begin{eqnarray}
\label{extended gauge symmetry tx}
{\cal A} &\to & {\cal A} + 2b \nonumber \\
{\bf g} & \to & {\bf g} + 2db,
\end{eqnarray}
where $b$ is a connection one-form on a $U(1)$-bundle $\cal N$ with curvature $db$, such that the usual Dirac quantisation condition $\int_{U} db/2\pi \in \mathbb Z$ is obeyed. Notice that with the above definition of an extended gauge symmetry, the structure of (\ref{Dirac quantisation})---relevant to an $SO(3)$ gauge theory---will be preserved as required; under (\ref{extended gauge symmetry tx}), we have $k_\gamma \to k_\gamma + n$ where $n = \int_{U_\gamma} db/2\pi$ is an integer, so $F$ is invariant modulo $4\pi$.

Two points to note before we proceed further are the following. The first point is that if $\cal N$ has trivial curvature with $2b = -{d\epsilon}$, one gets back an ordinary abelian gauge symmetry. Since $\cal A$ is supposed to be a connection on the bundle $\cal L$, it will mean that ${\cal A} + 2b$ must be a connection on the bundle ${\cal L} \otimes {\cal N}^{2}$. For trivial (or flat) $\cal N$, where one just has an ordinary abelian gauge theory, it is clear that it suffices to consider only some $\cal L$ in order to define the theory properly. However, in order to generalise the theory to nontrivial $\cal N$---that is, for $({\cal A} +2b)$ and $({\bf{g}} + 2db)$ to be physically valid as a gauge field and two-form, respectively---one must necessarily sum over all $\cal L$'s. The second point is that a consequence of an invariance of $\it{any}$ theory under (\ref{extended gauge symmetry tx}) is that one is free to shift the periods of $\bf g$---that is, the integrals of $\bf {g}$ over $U$---by integer multiples of $4 \pi$:
\be
\int_U {\bf{g}} \to \int_U {\bf{g}} + 4 \pi m, \quad \forall m \in \mathbb Z.
\label{g period}
\ee
(Again, we have made use of the fact that we have $\int_U db \in 2\pi \mathbb Z$).

\vspace{0.2cm}
\hspace{-1.0cm}(ii). \emph{The Corresponding Extended Theory}

Next, note that one way to modify ${\widehat I}'_{gauge}$ so that we can have invariance under the transformations (\ref{extended gauge symmetry tx}), is to replace $F'$ with ${\cal F}' = F' - \bf{g}$. However, notice that the resulting theory is trivial and not equivalent to the original theory, because one cannot set $\bf g$ to zero even if we let $2b = - d\epsilon$. Nevertheless, one can introduce a $\it{dual}$ abelian gauge field $\bf{w}$ (whose ``dual'' label will be justified shortly), that is a connection one-form on a $\it{dual}$ $U(1)$-bundle ${\cal L}_D$ with curvature ${\bf{W}} = d\bf{w}$, and add to the action ${\widehat I}'_{gauge}$ the term
\be
{\widetilde {\bf{I}}} = -{i \over 8 \pi} \int_M d^4 x \sqrt{h} \epsilon^{mnpq}{\bf W}_{mn} {\bf g}_{pq} = -{i \over 2 \pi} \int_M {\bf W} \wedge {\bf g}.
\ee
Like any curvature of an ordinary line bundle, we have the condition $\int_U {\bf W}/ 2 \pi \in \mathbb Z$.\footnote{Even though $\bf W$ will turn out to be the dual field strength $F_D$, it will not be constrained by the refinement (\ref{Dirac quantisation}) like $F$ would, because $F_D$ is $\it{not}$ the field strength of a $U(1)$ symmetry that is left unbroken from a non-abelian gauge symmetry that is supposedly dual to $SO(3)$ at high energies.} Thus, we find that ${\widetilde {\bf{I}}}$ is invariant modulo $4 \pi i \mathbb Z$ under the extended gauge transformation (\ref{extended gauge symmetry tx}). It is also invariant under the gauge transformation ${\bf w} \to {\bf w} - d {\widetilde \epsilon}$, where $\widetilde \epsilon$ is a zero-form on $M$.

Let us now define an extended theory in the fields $({\cal A}, {\bf g}, {\bf w})$ with action
\be
({\widetilde {\bf{I}}} + {\widehat I}'_{gauge}) ({\cal A}, {\bf g}, {\bf w}) = -{i \over 2 \pi} \int_M {\bf W} \wedge {\bf g} + {{i \bar\tau} \over 4\pi} \int_M {\cal F'}^{+} \wedge \star {\cal F'}^{+} - {{i \tau} \over 4\pi} \int_M {\cal F'}^{-} \wedge \star {\cal F'}^{-}  - i \eta \int_D {\cal F}'.
\label{extended action}
\ee
Since under (\ref{extended gauge symmetry tx}), ${\cal F}'$ is manifestly invariant while $\widetilde {\bf I}$ is invariant modulo $4 \pi i \mathbb Z$, we find that $({\widetilde {\bf{I}}} + {\widehat I}'_{gauge}) ({\cal A}, {\bf g}, {\bf w})$ will be invariant modulo $4 \pi i \mathbb Z$ under (\ref{extended gauge symmetry tx}), as required. It is also invariant under gauge transformations of ${\bf w}$.

\vspace{0.2cm}
\hspace{-1.0cm}(iii). \emph{The Equivalence Between the Extended and Original Theories}.

We would now like to show that the extended theory with action $({\widetilde {\bf{I}}} + {\widehat I}'_{gauge}) ({\cal A}, {\bf g}, {\bf w})$ is equivalent to the original theory with action ${\widehat I}'_{gauge}$. To this end, first note that the partition function of the extended theory can be written as
\be
{1 \over {\textrm{vol}(\cal G)}}{1 \over {\textrm{vol}(\widehat{\cal G})}}{1 \over {\textrm{vol}({\cal G}_D)}} \sum_{{\cal L},{{\cal L}_D}} \int {\cal D}{\cal A} \ {\cal D}{\bf g} \ {\cal D} {\bf w} \ \textrm{exp} \left(-({\widetilde {\bf{I}}} + {\widehat I}'_{gauge}) ({\cal A}, {\bf g}, {\bf w})\right),
\label{partition function}
\ee
where $\cal G$ and its $\it{dual}$ ${\cal G}_D$ denote the group of gauge transformations associated to $\cal A$ and $\bf{w}$, and $\widehat{\cal G}$ denotes the group of extended gauge transformations associated to $\bf g$. Next, let us try to compute the path-integral over the $\bf{w}$ fields. To do this, first write  ${\bf w} = {\bf w}_0 + {\bf w}'$, where ${\bf w}_0$ is a fixed connection on the line bundle ${\cal L}_D$. Then, the path-integral over the $\bf w$ fields can be written as
\be
{1 \over {\textrm{vol}({\cal G}_D)}} \sum_{{\cal L}_D} \int {\cal D}{\bf w}'  \ \textrm{exp}\left(  {i \over 2 \pi} \int_M {\bf w}' \wedge d{\bf g}\right) \cdot \textrm{exp}\left(  {i \over 2 \pi} \int_M {\bf W}_0 \wedge {\bf g}\right),
\label{path integral of w}
\ee
where ${\bf W}_0 = d {\bf w}_0$ corresponds to the curvature of the fixed connection ${\bf w}_0$. ${\bf W}_0$ is a closed two-form on $M$ in the cohomology $H^2(M)$, as ${\bf w}_0$ is only defined locally as a one-form. Noting that
\be
{1 \over {\textrm{vol}({\cal G}_D)}} \int {\cal D}{\bf w}'  \ \textrm{exp}\left(  {i \over 2 \pi} \int_M {\bf w}' \wedge d{\bf g}\right) = \delta (d {\bf g}),
\ee
one can compute (\ref{path integral of w}) as
\be
 \sum_{{\bf W}_0 \in H^2(M)} \textrm{exp}\left(  i \int_M {\bf W}_0 \wedge {{\bf g} \over 2 \pi}\right) \cdot \delta (d {\bf g}) = \delta \left( \left[{{\bf g} \over 2 \pi}\right] \in \mathbb Z \right) \cdot \delta (d {\bf g}).
\ee
In other words, we have the condition $d {\bf g} = 0$. We also have the condition that $\left[{{\bf{g}} \over 2 \pi}\right]$ belongs to an integral class, that is, the periods $\int_U {\bf{g}} \in 2 \pi \mathbb Z$. The first condition says that one can pick $\bf{g}$ to be a constant two-form. Together with the second condition and (\ref{g period}), one can indeed obtain ${\bf g} = 0$ via the extended gauge transformation (\ref{extended gauge symmetry tx}). By setting ${\bf g} = 0$,  the action (\ref{extended action}) will reduce to the original action (\ref{action gauge field}). Hence, the theory with action $({\widetilde {\bf{I}}} + {\widehat I}'_{gauge}) ({\cal A}, {\bf g}, {\bf w})$ is indeed equivalent to the original theory with action ${\widehat I}'_{gauge}$.

\vspace{0.2cm}
\hspace{-1.0cm}(iv). \emph{The Abelian Theory in the Dual, ``Magnetic'' Frame}.

Finally, we would like to ascertain ${\widehat I}'_{gauge}$ in the dual, ``magnetic'' frame. To this end, we shall make use of the equivalent action $({\widetilde {\bf{I}}} + {\widehat I}'_{gauge}) ({\cal A}, {\bf g}, {\bf w})$.

According to~\cite{SO abelian}, one ought to use the extended gauge symmetry (\ref{extended gauge symmetry tx}) to set ${\cal A} = 0$ for this purpose. However, note that because of the condition (\ref{Dirac quantisation}), and the fact that (\ref{extended gauge symmetry tx}) is only good enough to effect integral shifts in $k_\gamma$ as seen earlier, we conclude that for $c_\gamma \neq 0$ (that is, $w_2(E) \neq 0$), we cannot set ${\cal A} = 0$ using (\ref{extended gauge symmetry tx}). The best that we can do is to use (\ref{extended gauge symmetry tx}) to set $k_\gamma = 0$, so that $\int_{U_\gamma} F = 2 \pi c_\gamma$ (from (\ref{Dirac quantisation})), such that for a fixed set of nontrivial line bundles ${\cal L}_\gamma$ with connections $\zeta_{\gamma}$ and curvatures $G_\gamma$ in the corresponding cohomology classes, whereby
\be
\int_{U_\gamma} G_\beta = 2 \pi \delta_{\gamma\beta},
\label{fixed set}
\ee
we will have ${\cal A} = \sum_\gamma c_\gamma \zeta_\gamma$. Then, by shifting ${\bf g} \to {\bf g} + \sum_\gamma c_\gamma G_\gamma$ at the same time,\footnote{Note that we have implicitly assumed $M$ to be simply-connected in this instance, such that one can use the ordinary abelian gauge transformation to first set $\cal A$ to pure gauge (whilst preserving (\ref{Dirac quantisation})), before using (\ref{extended gauge symmetry tx}) to set ${\cal A} = \sum_\gamma c_\gamma \zeta_\gamma$ whilst effecting the shift ${\bf g} \to {\bf g} + \sum_\gamma c_\gamma G_\gamma$ simultaneously.} the extended action becomes
\begin{eqnarray}
({\widetilde {\bf{I}}} + {\widehat I}'_{gauge}) ({\bf g}, {\bf w}) & = & -{i \over 2 \pi} \int_M {\bf W} \wedge {\bf g} + {{i \bar\tau} \over 4\pi} \int_M {\bf g'}^{+} \wedge \star {\bf g'}^{+} - {{i \tau} \over 4\pi} \int_M {\bf g'}^{-} \wedge \star {\bf g'}^{-}  + i \eta \int_D {\bf g}' \nonumber \\
 && \quad -{i \over 2 \pi} \sum_\gamma  c_\gamma \int_M  {\bf W} \wedge G_\gamma,
\end{eqnarray}
where ${\bf g}' = {\bf g} + 2\pi \alpha \delta_D$.

Noting that
\be
\int_M {\bf W} \wedge {\bf g} = \int_M \left ( {\bf W}^+ \wedge \star {\bf g}^+ - {\bf W}^- \wedge \star {\bf g}^-      \right) = \int_M ({\bf W}^+ \cdot {\bf g}^+) - ({\bf W}^- \cdot {\bf g}^-),
\label{W identity}
\ee
one can rewrite the action as
\begin{eqnarray}
({\widetilde {\bf{I}}} + {\widehat I}'_{gauge}) ({\bf g}, {\bf w}) & = & - {i \over 2 \pi} \int_M ({\bf W}^+ - 2 \pi  \eta \delta^+_D) \cdot {\bf g}^+ - ({\bf W}^- - 2 \pi  \eta \delta^-_D) \cdot {\bf g}^- + {{i \bar\tau} \over 4\pi} \int_M |2 \pi \alpha \delta^+_D + {\bf g}^{+}|^2 \nonumber \\
&& - {{i \tau} \over 4\pi} \int_M |2 \pi \alpha \delta^-_D + {\bf g}^{-}|^2 -{i \over 2 \pi} \sum_\gamma  c_\gamma \int_M  {\bf W} \wedge G_\gamma,
\label{2.20}
\end{eqnarray}
where $|k|^2 = k \wedge \star k$ for any two-form $k$. Note that in the above, we have also used the fact that the term $2 \pi i \eta \alpha \int_M \delta_D \wedge \delta_D$ which appears in the action can be set to zero modulo $2\pi i \mathbb Z$; recall that we are either considering trivially-embedded surface operators for which $\int_M \delta_D \wedge \delta_D = D \cap D = 0$, or nontrivially-embedded surface operators whereby $\alpha D\cap D$ and $\eta$ are both integers. (The appearance of this term is indeed consistent with the observation in (\ref{crucial diff}) of the full supersymmetric theory.)

If we define
\be
{\bf g}' = {\bf g} + 2 \pi \alpha \delta_D - {1 \over {\bar\tau}} \left ({\bf W}^+ - 2 \pi \eta \delta^+_D \right) -  {1 \over {\tau}} \left ( {\bf W}^- - 2 \pi \eta \delta^-_D \right),
\label{g'}
\ee
we can rewrite the action as
\begin{eqnarray}
\label{I g'}
 ({\widetilde {\bf{I}}} + {\widehat I}'_{gauge}) ({\bf g}, {\bf w}) & = & {{i \bar\tau} \over {4 \pi}} \int_M |{\bf g}^{'+}|^2 - {{i \tau} \over {4 \pi}} \int_M |{\bf g}^{'-}|^2  - {i \over {4 \pi \bar\tau}} \int_M |{\bf W}^+ - 2 \pi \eta \delta^+_D|^2  \\
&& + {i \over {4 \pi \tau}} \int_M |{\bf W}^- - 2 \pi \eta \delta^-_D|^2  + i \alpha \int_D ({\bf W} - 2 \pi \eta \delta_D) -{i \over 2 \pi} \sum_\gamma  c_\gamma \int_M  {\bf W} \wedge G_\gamma.\nonumber
\end{eqnarray}
Then, by integrating out the ${\bf g}^{'+}$ and ${\bf g}^{'-}$ fields classically using the Euler-Lagrange equations, we have
\begin{eqnarray}
 ({\widetilde {\bf{I}}} + {\widehat I}'_{gauge}) ({\bf w})& = &  -{i \over {4 \pi \bar\tau}} \int_M {\bf W'}^+  \wedge \star {\bf W'}^+ + {i \over {4 \pi \tau}} \int_M {\bf W'}^- \wedge \star {\bf W'}^- + i \alpha \int_D {\bf W'} \nonumber \\
&& -{i \over 2 \pi} \sum_\gamma  c_\gamma \int_M  {\bf W} \wedge G_\gamma,
\label{2.23}
\end{eqnarray}
where ${\bf W'} = {\bf W} - 2 \pi \eta \delta_D$. Comparing this with (\ref{action gauge field}), we have the following physical equivalence of actions
\be
{\widehat I}'_{gauge}({\cal A}, \tau, \alpha, \eta) \equiv {\widehat I}'_{gauge}({\cal A}_D, \tau_D, \alpha_D, \eta_D) -{i \over 2 \pi} \sum_\gamma  c_\gamma \int_M  F_D \wedge G_\gamma.
\label{relation}
\ee
where $\tau_D = -1/\tau$, $\alpha_D = \eta$ and $\eta_D = - \alpha$, as expected, and ${\cal A}_D$ is the gauge field for the ``magnetic'' photon.

\vspace{0.2cm}
\smallskip\noindent{\it The Dependence on $w_2(E)$}

From (\ref{relation}), we learn that for the theory in the ``magnetic'' frame near $u=1$, one must include in the path-integral the additional phase factor
\be
\textrm{exp}\left({i \over 2 \pi} \sum_\gamma  c_\gamma \int_M  F_D \wedge G_\gamma\right).
\ee
From (\ref{stiefel-whitney}) and (\ref{fixed set}), we find that we can write
\be
w_2(E) = \sum_\gamma c_\gamma \left[{G_\gamma \over 2\pi}\right].
\label{w_2(E)}
\ee
%where $ \left[{G_\gamma / 2\pi}\right]$ is the cohomology class of the curvature two-form ${G_\gamma / 2\pi}$.
Consequently, the phase factor will be given by
\be
(-1)^{(c_1({\cal L}^{\otimes 2}_D), w_2(E))},
\label{phase 2}
\ee
where
\be
(c_1({\cal L}^{\otimes 2}_D), w_2(E)) = \int_M c_1({\cal L}^{\otimes 2}_D)\wedge w_2(E),
\ee
and $c_1({\cal L}^{\otimes 2}_D) = 2 F_D / 2\pi$.

Alternatively, if we were to trivially re-scale (\ref{relation}) by an overall factor of $1/2$ (so as to agree with the definition of the action in~\cite{abelian S-duality}), the phase factor would be given by
\be
(-1)^{(c_1({\cal L}_D), w_2(E))}.
\label{phase 2-alternative}
\ee
Hence, we see a dependence on $w_2(E)$ via this additional phase factor that must be included in the path-integral whenever one is dealing with an $SO(3)$ gauge theory that has $w_2(E) \neq 0$. The result of (\ref{phase 2-alternative}) is again exactly the same as that found  in~\cite{abelian S-duality} for the case without surface operators. This should perhaps not be so surprising. After all, the description of the gauge field strengths as characteristic classes does not make any reference to the explicit form of the gauge connections. Consequently, their topological properties and hence the result of (\ref{phase 2}) should not be modified in the presence of  ``ramification''. One only has to be careful when evaluating an integral of a ``ramified'' field strength over some region in $M$ that contains $D$, and our above arguments have not, at any point, required us to do so.

\newsubsection{Appearance of a $\textrm{Spin}^c$ Structure and an Interesting Phase Factor}

In all our discussions in $\S$3 and $\S$4 so far, we have implicitly assumed that the manifold is spin, that is, $w_2(M) = 0$. As such, additional interaction terms that might appear when $M$ is non-spin have yet to be considered. In the case without surface operators, such a term has been explicitly determined in~\cite{abelian S-duality}; where it has been shown to arise when one integrates out the massive fermions that are the $SU(2)$ (or $SO(3)$) partners of $\lambda$ and $\bar\lambda$.

Note that since the analysis in~\cite{abelian S-duality} leading to the determination of the interaction term only involves the topological description of gauge field strengths in terms of characteristic classes (that make no reference to the explicit form of the gauge connection), the result would be similar even when one includes surface operators. In particular, this means that one ought to have the following additional interaction term in the action when $M$ is non-spin:
\be
{\widehat I}'_{int} = {i \over 4 \pi} \sum_\gamma e^\gamma \int_M  F' \wedge G_\gamma,
\label{result}
\ee
where $e^\gamma$ are integers and
\be
w_2(M) \equiv \sum_\gamma e^\gamma \left[{G_\gamma \over 2 \pi}\right] \ \textrm{mod} \ 2.
\label{egamma condition}
\ee
Notice that we have again replaced $F$ with $F'$ in the result from~\cite{abelian S-duality} in writing (\ref{result}) above, as we are now including surface operators in the theory.

Clearly, (\ref{result}) vanishes when $M$ is spin. However, when $w_2(M) \neq 0$, a $\textrm{Spin}^c$ structure in the ``magnetic'' theory near $u=1$ would make an appearance because of (\ref{result}). We shall now demonstrate this claim.

\vspace{0.0cm}
\smallskip\noindent{\it Abelian Duality in Low-Energy $SU(2)$ Gauge Theory}

In order to do so, it suffices again to analyse only the gauge field dependent part of the action. As before, since the analysis will be independent of the explicit form of $\tau$, we shall take ${\widehat I}'_{gauge}$ to be the gauge field dependent part of the topological action ${\widehat I}'$ of the abelian ``electric'' theory. Also, it would be more illuminating to consider, at this point, the case where $w_2(E)= c_\gamma = 0$. Hence, any $SO(3)$-bundle can be lifted to an $SU(2)$-bundle, and the condition (\ref{c1}) which leads to (\ref{Dirac quantisation}) will no longer hold. So let us consider the microscopic theory to be an $SU(2)$ gauge theory; in other words, the low-energy ``electric'' abelian gauge field will be $C$, where $F = dC$ and $F/2\pi = c_1(T)$.

The effective action ${\widehat I}'_{\textrm{eff}}$ to consider would be ${\widehat I}'_{gauge}(C, \tau, \alpha, \eta) + {\widehat I}'_{int}$, that is,
\be
{\widehat I}'_{\textrm{eff}} = {i\bar\tau\over 4 \pi} \int_M {F'}^+ \wedge \star {F'}^+ - {i\tau\over 4 \pi} \int_M {F'}^- \wedge \star {F'}^- - i\eta \int_D F' + {i \over 4 \pi} \sum_\gamma e^\gamma \int_M  F' \wedge G_\gamma.
\label{I_eff}
\ee
The analysis is pretty much the same as before, except for a few differences. Let us comment on them while we proceed with our computation.

\vspace{0.2cm}
\hspace{-1.0cm}(i). \emph{An Extended Gauge Symmetry}

Firstly, because of the absence of the refined Dirac quantisation condition (\ref{Dirac quantisation}), the extended gauge symmetry is now
\begin{eqnarray}
\label{extended gauge symmetry tx SU(2)}
{C} &\to & {C} + b \nonumber \\
{\bf g} & \to & {\bf g} + db,
\end{eqnarray}
without the factor of two. Consequently, under (\ref{extended gauge symmetry tx SU(2)}), the periods of $\bf g$ would  be shifted by
\be
\int_U {\bf g} \to \int_U {\bf g} + 2 \pi m, \quad \forall m \in \mathbb Z.
\label{g period SU(2)}
\ee

\vspace{0.2cm}
\hspace{-1.0cm}(ii). \emph{The Corresponding Extended Theory}

The extended theory, with the additional interaction, now reads
\begin{eqnarray}
({\widetilde {\bf{I}}} + {\widehat I}'_{\textrm{eff}}) ({C}, {\bf g}, {\bf w}) & = & -{i \over 2 \pi} \int_M {\bf W} \wedge {\bf g} + {{i \bar\tau} \over 4\pi} \int_M {\cal F'}^{+} \wedge \star {\cal F'}^{+} - {{i \tau} \over 4\pi} \int_M {\cal F'}^{-} \wedge \star {\cal F'}^{-} \nonumber \\
 && - i \eta \int_D {\cal F}' + {i \over 4 \pi} \sum_\gamma e^\gamma \int_M  {\cal F}' \wedge G_\gamma.
\label{extended action SU(2)}
\end{eqnarray}
As required, since ${\cal F}' = F' - {\bf g}$ is manifestly invariant under (\ref{extended gauge symmetry tx SU(2)}), while $\widetilde I$ is now invariant modulo $2\pi i \mathbb Z$ under (\ref{extended gauge symmetry tx SU(2)}) (because $\int_U {\bf W} /2\pi \in \mathbb Z$), the extended action is invariant modulo $2\pi i \mathbb Z$ under (\ref{extended gauge symmetry tx SU(2)}), as required.

\vspace{0.2cm}
\hspace{-1.0cm}(iii). \emph{The Equivalence Between the Extended and Original Theories}.

The partition function of the extended theory will, in this case, be expressed in terms of the gauge field $C$ and its $SU(2)$-bundle $T$:
\be
{1 \over {\textrm{vol}(\cal G)}}{1 \over {\textrm{vol}(\widehat{\cal G})}}{1 \over {\textrm{vol}({\cal G}_D)}} \sum_{{T},{{T}_D}} \int {\cal D}{C} \ {\cal D}{\bf g} \ {\cal D} {\bf w} \ \textrm{exp} \left(-({\widetilde {\bf I}} + {\widehat I}'_{\textrm{eff}}) ({C}, {\bf g}, {\bf w})\right),
\label{partition function}
\ee
where $\cal G$ and its $\it{dual}$ ${\cal G}_D$ denote the group of gauge transformations associated to $C$ and $\bf{w}$, and $\widehat{\cal G}$ denotes the group of extended gauge transformations associated to $\bf g$.

Similarly, one can set $\bf g$ to zero via the condition $d {\bf g} = 0$ and (\ref{g period SU(2)}), like before. As a result, the extended theory with action ${\widetilde {\bf I}} + {\widehat I}'_{\textrm{eff}}$ is equivalent to the original theory with action ${\widehat I}'_{\textrm{eff}}$.

\vspace{0.2cm}
\hspace{-1.0cm}(iv). \emph{The Abelian Theory in the Dual, ``Magnetic'' Frame}.

We are now ready to ascertain ${\widehat I}'_{\textrm{eff}}$ in the dual, ``magnetic'' frame. To this end, we shall make use of the equivalent action $({\widetilde {\bf{I}}} + {\widehat I}'_{\textrm{eff}}) ({C}, {\bf g}, {\bf w})$.

Since we are not constrained by the condition (\ref{Dirac quantisation}), we can use the extended gauge symmetry (\ref{extended gauge symmetry tx SU(2)}) to set ${C} = 0$. The extended action then becomes
\begin{eqnarray}
({\widetilde {\bf{I}}} + {\widehat I}'_{\textrm{eff}}) ({\bf g}, {\bf w}) & = & -{i \over 2 \pi} \int_M {\bf W} \wedge {\bf g} + {{i \bar\tau} \over 4\pi} \int_M {\bf g'}^{+} \wedge \star {\bf g'}^{+} - {{i \tau} \over 4\pi} \int_M {\bf g'}^{-} \wedge \star {\bf g'}^{-}  + i \eta \int_D {\bf g}' \nonumber \\
 && \quad -{i \over 4 \pi} \sum_\gamma e^\gamma \int_M  {\bf g}' \wedge G_\gamma,
\end{eqnarray}
where ${\bf g}' = {\bf g} + 2\pi \alpha \delta_D$.

Let us define
\be
{\bf w}^c = {\bf w} + {1\over 2} \sum_\gamma e^\gamma \zeta_\gamma.
\label{spin^c condition}
\ee
Then, we can rewrite the extended action as
\begin{eqnarray}
({\widetilde {\bf{I}}} + {\widehat I}'_{\textrm{eff}}) ({\bf g}, {\bf w}^c) & = & -{i \over 2 \pi} \int_M {\bf W}^c \wedge {\bf g} + {{i \bar\tau} \over 4\pi} \int_M {\bf g'}^{+} \wedge \star {\bf g'}^{+} - {{i \tau} \over 4\pi} \int_M {\bf g'}^{-} \wedge \star {\bf g'}^{-} + i \eta \int_D {\bf g}' \nonumber \\
&&  - {i \alpha \over 2} \sum_\gamma e^\gamma \int_D G_\gamma,
\end{eqnarray}
where ${\bf W}^c = d {\bf w}^c$.

We can repeat the computation in (\ref{W identity})-(\ref{2.23}), replacing $\bf W$ with ${\bf W}^c$ and the last term in (\ref{2.20}) with $- {i \alpha \over 2} \sum_\gamma e^\gamma \int_D G_\gamma$, and we find that
\begin{eqnarray}
 ({\widetilde {\bf{I}}} + {\widehat I}'_{\textrm{eff}}) ({\bf w}^c)& = &  -{i \over {4 \pi \bar\tau}} \int_M {\bf W'}^{c+}  \wedge \star {\bf W'}^{c+} + {i \over {4 \pi \tau}} \int_M {\bf W'}^{c-} \wedge \star {\bf W'}^{c-} + i \alpha \int_D {\bf W'}^c \nonumber \\
&& - {i \alpha \over 2} \sum_\gamma e^\gamma \int_D G_\gamma,
\label{2.23 SU(2)}
\end{eqnarray}
where ${\bf W'}^c  = {\bf W}^c - 2\pi \eta \delta_D$. Comparing this with (\ref{I_eff}), we have the following physical equivalence of actions
\be
{\widehat I}'_{\textrm{eff}}({C}, \tau, \alpha, \eta, e^\gamma, \zeta_\gamma) \equiv {\widehat I}'_{gauge}(C^c_D, \tau_D, \alpha_D, \eta_D) - {i \alpha \over 2} \sum_\gamma e^\gamma \int_D G_\gamma,
\label{relation SU(2)}
\ee
where $\tau_D = -1/\tau$, $\alpha_D = \eta$ and $\eta_D = - \alpha$, while $C^c_D$ is the gauge field for the ``magnetic'' photon.

\vspace{0.2cm}
\smallskip\noindent{\it The Appearance of a $\textrm{Spin}^c$ Structure}

Notice that the dual field strength $F^c_D = dC^c_D$ associated to the ``magnetic'' theory near $u=1$ is such that
\be
F^c_D = F_D + {1\over 2} \sum_\gamma e^\gamma G_\gamma,
\label{F^C_D}
\ee
where $F_D$ is the dual ``magnetic'' field strength that one would have had instead if the additional interaction ${\widehat I}'_{int}$ was absent. Even though $F_D$ and $G_\gamma$ are integral classes, the presence of $1/2$ in (\ref{F^C_D}) means that $F^c_D$ $\it{cannot}$ be an integral class, that is, it will not correspond to a curvature of an ordinary complex line bundle.

The only way that $F^c_D$ can be an integral class is when the $e^\gamma$'s are all even integers. However, when this is so, we see from (\ref{egamma condition}) that $w_2(M)$ will effectively vanish. In other words, the obstruction to $F^c_D$ being a curvature of an ordinary complex line bundle is $w_2(M)$. In turn, this implies that there must exist a bundle ${\cal T} = T_D^{c\otimes 2}$ (where $F^c_D$ is the curvature of the ``bundle'' $T^c_D$), such that $c_1({\cal T}) \in 2 H^2(M, \mathbb Z) + w_2(M)$ (since (\ref{egamma condition}) means that $w_2(M)$ is an integral class). Such a bundle $\cal T$ is known as a determinant line bundle of a $\textrm{Spin}^c$ structure~\cite{Marcos}. Consequently, this means that $T^c_D$ cannot exist as a bundle by itself, but the product $S_+ \otimes T^c_D$, where $S_+$ is a positive chirality spinor bundle on $M$, does exist, and is termed a $\textrm{Spin}^c$ structure. Indeed, the monopole field $\cal M$ of the Seiberg-Witten equations that appeared in footnote~8, will, in this case, be a section of $S_+ \otimes T^{c'}_D$,  and it is always physically well-defined.

\vspace{0.2cm}
\smallskip\noindent{\it An Interesting Phase Factor}

From the relation (\ref{relation SU(2)}), we find that one must include the following phase factor in the Euclidean path-integral near $u=1$:
\be
\textrm{exp} \left(  {i \alpha \over 2} \sum_\gamma e^\gamma \int_D G_\gamma \right).
\ee
Recall our assumption that there is no torsion in the cohomology of $M$, and that $D$ has been defined to be closed and oriented. If we further assume that $D$ is not some boundary of a three-manifold in $M$,  we can expand $D$ as
\be
D  = \sum_\beta d^\beta U_\beta,
\ee
where the $U_\beta$'s are a basis of two-homology cycles in $H_2(M,\mathbb Z)$ such that the $d^\beta$'s are integers. Then, the condition on the $G_\gamma$'s in (\ref{fixed set}) will imply that one can write the phase factor as
\be
\left( -1 \right)^{\alpha ({e \cdot d})},
\label{p}
\ee
where
\be
(e \cdot d) = \sum_\gamma e^\gamma d^\gamma.
\ee
If we were to trivially re-scale the relation (\ref{relation SU(2)}) by an overall factor of $1/2$ (so as to agree with the definition of the action in~\cite{abelian S-duality}), the exponent of the phase factor---${\alpha ({e \cdot d})}$---would just be multiplied by $1/2$.

Thus, we see a dependence on the ``classical'' surface operator parameter $\alpha$ in a phase factor of the $\it{quantum}$ path-integral of the dual ``magnetic'' theory. This observation is consistent with the fact that $S$-duality maps a ``classical'' parameter to a ``quantum'' parameter and vice-versa (the prototype relations being $\alpha_D = \eta$ and $\eta_D = - \alpha$, where $\eta$ is a ``quantum'' parameter).

Note that the phase factor (\ref{p}) is not necessarily trivial even when $M$ is spin; recall that it suffices for the $e^\gamma$'s to be even integers for $w_2(M)$ to effectively vanish (and thus, $M$ to be spin), but the exponent ${\alpha ({e \cdot d})}$ in (\ref{p}) is zero if and only if $\alpha$ and/or $(e \cdot d)$ vanish.

Last but not least, as will be elaborated in forthcoming work~\cite{MC}, the phase factor (\ref{p}) is also a crucial ingredient in the physical proof of the relation between the ``ramified'' and ``unramified'' Donaldson invariants, as first established by Kronheimer and Mrowka in~\cite{KM}.

\newsubsection{Combining the Two Effects}

Finally, it will be useful for future computations to derive the combined effect of having $w_2(E) \neq 0$ and $w_2(M) \neq 0$.

Firstly, $w_2(E) \neq 0$ means that we will necessarily have to consider an $SO(3)$ gauge theory at high energies. In addition, one is also subject to the refined Dirac quantisation condition (\ref{Dirac quantisation}). As such, it would mean that we can never use the extended gauge invariance to set ${\cal A} = 0$. However, we can still set ${\cal A} = \sum_\gamma c_\gamma$ whilst shifting ${\bf g} \to {\bf g} + \sum_\gamma c_\gamma G_\gamma$, as explained earlier. Secondly, $w_2(M) \neq 0$ means that the interaction term (\ref{result}) cannot be ignored. Altogether, the extended action is now given by
\begin{eqnarray}
{\widehat I}'_{\textrm{ext}} ({\bf g}, {\bf w}) & = & -{i \over 2 \pi} \int_M {\bf W}^c \wedge {\bf g} + {{i \bar\tau} \over 4\pi} \int_M {\bf g'}^{+} \wedge \star {\bf g'}^{+} - {{i \tau} \over 4\pi} \int_M {\bf g'}^{-} \wedge \star {\bf g'}^{-}  + i \eta \int_D {\bf g}'  \nonumber \\
 && \quad -{i \over 2 \pi} \sum_\gamma  c_\gamma \int_M  {\bf W} \wedge G_\gamma - {i\alpha \over 2} \sum_\gamma e^\gamma \int_D G_\gamma.
\end{eqnarray}
%where ${\bf g}' = {\bf g} + 2\pi \alpha \delta_D$, ${\bf W}^c = d{\bf w}^c$, and ${\bf w}^c = {\bf w} + {1\over 2} \sum_\gamma e^\gamma \zeta_\gamma$.
Integrating out ${\bf g}$ like we did before, we get
\begin{eqnarray}
{\widehat I}'_{\textrm{ext}} ({\cal A}^c_D, \tau_D, \alpha_D, \eta_D)& = &  {i \bar\tau_D \over {4 \pi}} \int_M ({F^{c'}_D})^{+}  \wedge \star ({F^{c'}_D})^{+} - {i \tau_D \over {4 \pi}} \int_M ({F^{c'}_D})^{-} \wedge \star ({F^{c'}_D})^{-} - i \eta_D \int_D {F^{c'}_D} \nonumber \\
&&   -{i \over 4 \pi} \sum_\gamma  c_\gamma \int_M  \left(2F^c_D - \sum_\beta e^\beta G_\beta\right) \wedge G_\gamma  - {i \alpha \over 2} \sum_\gamma e^\gamma \int_D G_\gamma.
\label{Icom}
\end{eqnarray}
In other words, if we denote the original combined action as ${\widehat I}'_{\textrm{com}}$, we would have the following physical equivalence of actions:
\be
{\widehat I}'_{\textrm{com}} ({\cal A}, \tau, \alpha, \eta, e^\gamma, \zeta_\gamma) \equiv {\widehat I}'_{gauge}({\cal A}^c_D, \tau_D, \alpha_D, \eta_D) + \Delta {\widehat I}'_{\textrm{ext}},
\label{combo relation}
\ee
where
\be
\Delta {\widehat I}'_{\textrm{ext}} = -{i \over 4 \pi} \sum_\gamma  c_\gamma \int_M  \left(2F^c_D - \sum_\beta e^\beta G_\beta\right) \wedge G_\gamma  - {i \alpha \over 2} \sum_\gamma e^\gamma \int_D G_\gamma,
\ee
${F^{c}_D} = d {\cal A}^c_D$, and ${\cal A}^c_D = {\cal A}_D + {1\over 2} \sum_\gamma e^\gamma \zeta_\gamma$. As always, $\tau_D = - 1/\tau$, $\alpha_D = \eta$ and $\eta_D = - \alpha$.

Hence, because we have ${\cal A}^c_D$ and not simply ${\cal A}_D$, we find that near $u=1$, the ``magnetic'' theory would have a $\textrm{Spin}^c$ structure $S_+ \otimes {\cal L}^c_D$ (where ${\cal L}^c_D$ is the ``magnetic'' dual of the $SO(3)$-bundle $\cal L$), of which the monopole field $\cal M$ is a section of $S_+ \otimes {\cal L}^{c'}_D$. Moreover, one must also include the phase factor $\textrm{exp} ( - \Delta {\widehat I}'_{\textrm{ext}})$ in the path-integral when computing the contributions at $u = \pm 1$ (of the corresponding Seiberg-Witten invariants) to the generating function of the ``ramified'' Donaldson invariants.

From (\ref{w_2(E)}) and (\ref{egamma condition}), we gather that $w_2(E)$ and $w_2(M)$ are both integral classes. Also, note that $2F^c_D /2\pi = c_1({\cal L}_D^{c  \otimes 2})$, and since we have a $\textrm{Spin}^c$ structure, $c_1({\cal L}_D^{c  \otimes 2}) \in w_2(M) + 2 H^2(M,\mathbb Z)$; thus, $c_1({\cal L}_D^{c  \otimes 2})$ is an integral class too. Altogether, this means that we can write the $\it{combined}$ phase factor as
\be
\left( - 1 \right)^{ \Delta + \alpha (e\cdot d)},
\label{phase factor}
\ee
where
\be
\Delta = \left(c_1({\cal L}_D^{c \otimes 2}) -w_2(M) , w_2(E)\right).
\ee
Alternatively, if we were to re-scale the relation (\ref{combo relation}) by an overall factor of $1/2$ (so as to agree with the definition of the action in~\cite{abelian S-duality}), the exponent ${ \Delta + \alpha (e\cdot d)}$ in the phase factor (\ref{phase factor}) would just be multiplied by $1/2$.

A final observation which can be made is that the ``$\it{unramified}$'' part of (\ref{phase factor}) that survives as we let $\alpha, \eta \to 0$, can be expressed as
\be
\left( - 1 \right)^{\Delta} = e^{2i \pi ({\Lambda_0} , {\Lambda})}
\ee
modulo a factor of $(-1)^{{1\over 2}(w_2(E),w_2(M))}$, where $\Lambda_0 \in {1\over 2} w_2(E) + H^2(M,\mathbb Z)$ and $\Lambda = {1\over 2} c_1({\cal L}_D^{c  \otimes 2})$. This agrees exactly with the result in~\cite{Moore-Witten} for the combined phase factor of the theory without surface operators at $u= \pm 1$.

\vspace{1.0cm}
\hspace{-1.0cm}{\large \bf Acknowledgements:}\\
\vspace{-0.5cm}

I would like to thank S.~Gukov for illuminating discussions. This work is supported by the California Institute of Technology and the NUS-Overseas Postdoctoral Fellowship.

\end{document}